\documentclass[11pt]{article}
\usepackage[margin=1in]{geometry}

\usepackage[utf8]{inputenc}
\usepackage[english]{babel}
\usepackage{array}
\usepackage{amsmath, amssymb}
\usepackage{bbold}
\usepackage{amsthm} 
\usepackage{slashed}
\usepackage{mathtools} 
\usepackage{comment} 
\usepackage{setspace}
\usepackage[numbers,sort&compress]{natbib}
\usepackage{url}
\usepackage{footnote}
\usepackage[all]{xy}
\usepackage{tikz-cd, tikz-3dplot}
\usepackage{hyperref}
\usepackage{lscape}
\hypersetup{
    colorlinks,
    citecolor=black,
    filecolor=black,
    linkcolor=black,
    urlcolor=black
}
\usepackage{mathrsfs}
\usepackage{float}
\usepackage{subcaption}
\usetikzlibrary{positioning}

\DeclareMathOperator*{\argmax}{arg\,max}

\theoremstyle{definition}

\newcommand{\un}[1]{\underline{#1}}

\begin{document}


\begin{flushright}
\end{flushright}
\bigskip

\centerline{\Large{Bayesian Renormalization}}
\centerline{\Large{}}


\centerline{David S. Berman$^{1}$, Marc S. Klinger$^{2, *}$, Alexander G. Stapleton$^{1}$}

\centerline{${}^1$ Centre for Theoretical Physics, Queen Mary University of London, Mile End Road, London E1 4NS}
\centerline{${}^{2}$ Department of Physics, University of Illinois, Urbana IL 61801, USA}

\bigskip
\centerline{
    \vspace{0.3cm}
    {\tt \small
        $^*$Corresponding author: marck3@illinois.edu}}

\begin{abstract}
In this note we present a fully information theoretic approach to renormalization inspired by Bayesian statistical inference, which we refer to as Bayesian Renormalization. The main insight of Bayesian Renormalization is that the Fisher metric defines a correlation length that plays the role of an emergent RG scale quantifying the distinguishability between nearby points in the space of probability distributions. This RG scale can be interpreted as a proxy for the maximum number of unique observations that can be made about a given system during a statistical inference experiment. The role of the Bayesian Renormalization scheme is subsequently to prepare an effective model for a given system up to a precision which is bounded by the aforementioned scale. In applications of Bayesian Renormalization to physical systems, the emergent information theoretic scale is naturally identified with the maximum energy that can be probed by current experimental apparatus, and thus Bayesian Renormalization coincides with ordinary renormalization. However, Bayesian Renormalization is sufficiently general to apply even in circumstances in which an immediate physical scale is absent, and thus provides an ideal approach to renormalization in data science contexts. To this end, we provide insight into how the Bayesian Renormalization scheme relates to existing methods for data compression and data generation such as the information bottleneck and the diffusion learning paradigm. We conclude by designing an explicit form of Bayesian Renormalization inspired by Wilson's momentum shell renormalization scheme in Quantum Field Theory. We apply this Bayesian Renormalization scheme to a simple Neural Network and verify the sense in which it organizes the parameters of the model according to a hierarchy of information theoretic importance.  

\end{abstract}

\tableofcontents

\pagebreak

\section{Introduction}

Perhaps the central question in data science is the following: How does our understanding of a system improve as we obtain more data? The natural language for formulating this question is through statistical inference \cite{jaynes1957information, jaynes2003probability, gelman2013philosophy}. From the perspective of statistical inference, our understanding of a system is encoded in the probability we assign to different plausible explanations for how the system works. These explanations are formalized as probability models for observable data specified in terms of various parameters. The probability assigned to each of these models is subsequently encoded in an object called the Bayesian posterior distribution, which can be thought of as a probability distribution over all possible probability distributions for observable data. In terms of these concepts, \cite{berman2022dynamical} presents an answer to the aforementioned question by deriving an explicit equation governing the evolution of the posterior distribution as a function of the amount of collected data. We refer to this equation and more broadly to the idea of dynamically updating one's beliefs in light of new data using Bayesian inference as Dynamical Bayesian Inference, or Dynamical Bayes (DB). A central observation from DB is that as new data is collected the ``current" most likely model flows through the space of possible models towards the probability distribution truly responsible for generating observed data. 

The idea that learning induces a flow in the space of models is immediately quite evocative of a different kind of ``meta"-theory: the Renormalization Group (RG). RG is a set of ideas and strategies broadly concerned with formalizing the role of scale in our understanding and formulation of physical theories. In its original form, as conceived of by Kadanoff and Wilson \cite{kadanoff1966scaling,wilson1971renormalization,
wilson1974renormalization}, the renormalization group consists of taking a system described by a large number of degrees of freedom and performing a coarse-graining operation in which subsets of degrees of freedom are combined together and averaged over to form new collective variables. In physics applications, coarse-graining neighborhoods are determined based on considerations related to locality -- that is, degrees of freedom that are nearby in physical space are joined together. For this reason, the renormalization group takes a theory which describes behaviors of a system down to arbitrarily small scales to a new theory which describes behaviors only up to a distance scale that is constrained by the size of the typical coarse graining neighborhood. In the terminology of physics, we say that a renormalization group flow takes a UV theory (a theory that is valid at arbitrarily small scales, or equivalently arbitrarily high energies) to an IR theory (a theory that is valid only at relatively large distances, or equivalently relatively low energies). From the data science perspective, one may think of an IR theory as corresponding to a naive model in which a large amount of data, namely the fine grained description of the system at length scales smaller than the prescribed cutoff, have yet to be incorporated. By contrast, a UV theory has incorporated most if not all of the available data about the system and therefore corresponds to the complete data generating model or ``ground truth". In what follows we shall often make use of this interpretation of the UV and IR. For our purposes, we will be interested in a relatively modern incarnation of renormalization that typically goes under the name of the Exact Renormalization Group (ERG) \cite{wegner1974some, wegner1973renormalization, polchinski1984renormalization, morris1994derivative, morris1994exact, morris1998elements, latorre2000exact, bagnuls2001exact, Morris:2006in, rosten2012fundamentals}. ERG seeks to formalize the ideas of renormalization in a more mathematically rigorous fashion, by formulating an ERG flow as a one parameter family of theories governed by a (functional) differential equation. 

Viewing a physical theory as a probability distribution for whatever observable degrees of freedom make up the system, the effect of an RG coarse-graining scheme is therefore to induce a flow through the space of possible theories -- just like in DB. However, in contrast to the case of learning in which one flows through the space of models towards the data generating model, an RG flow begins at the data generating model (a UV complete theory) and flows away towards some less complete model which remains accurate only for a subset of the original degrees of freedom. This observation motivates the idea that the renormalization group flow may be regarded as a procedure ``inverse" to that of dynamical Bayes, with the former taking a data generating model down to an approximate model and the latter taking an approximate model back to the data generating model. This idea was formalized in \cite{berman2022inverse}, in which we showed that the equation governing dynamical Bayes is formally equivalent to an Exact Renormalization Group (ERG) flow if we invert the direction of the flow. More explicitly, performing dynamical Bayes \emph{in reverse} by discarding data as opposed to observing data \emph{defines} an ERG scheme which we refer to as the Dynamical Bayesian Renormalization Group scheme (DB-RG) or simply \emph{Bayesian Renormalization}. 

In this note we aim to flesh out Bayesian Renormalization. In particular, we would like to stress how the DB-RG scheme frees renormalization from its reliance on physical locality. As we alluded to above, when performing Renormalization for physical systems one implements a coarse-graining scheme that is directly motivated by physical locality e.g. either one defines collective variables by pooling together degrees of freedom that are contained in a common spatial neighborhood, or one integrates out degrees of freedom that have support on momentum shells above a particular high energy cutoff. In either case, the existence of a hierarchy of physical scales and their role in defining the RG scheme ensures that we can interpret an ERG flow as beginning from a UV theory and ending at some IR fixed point. But what are we to do if the physical system we are interested in renormalizing has non-local interactions? Or worse yet, what if we are interested in renormalizing a model that doesn't have a physical interpretation at all? Such a situation presents itself in recent work which seeks to import the machinery of renormalization into data science contexts as a tool for performing data compression and improving the interpretability and performance of high dimensional models \cite{Meshulam2018CoarseGF, Meshulam2018CoarsegrainingAH, Kline:2021ugg,mehta2014exact,Lin:2016fab, Halverson:2020trp, Luo:2021nab, Halverson:2021aot, Brown:2021rmz, He:2023csq, erdmenger2021quantifying}. 

Bayesian Renormalization overcomes the apparent lack of ``real" scale by coming equipped with its own \emph{emergent} scale -- the distinguishability of models. Put differently, the space of models has a natural information geometric stucture \cite{amari2000methods,amari2016information,nielsen2020elementary} endowed by the Fisher metric, which is an infinitesimal measure of the relative entropy between a pair of probability distributions. As we will demonstrate, the DB-RG scheme automatically coarse-grains in a way that respects locality in the space of models, as dictated by the Fisher metric. This fact is particularly serendipitous in touting the utility of DB-RG for renormalizing data science models. In data compression tasks and model building considerations the Fisher metric is used to distinguish between so-called ``sloppy" and ``stiff" parameters. The former covary only weakly with the model output and therefore correspond to small eigenvalues of the Fisher metric, while the latter covary very strongly with the model output and therefore correspond to large eigenvalues of the Fisher metric. To hone the interpretability and generalizability of a model, one may therefore be interested in a scheme which systematically discards sloppy parameters in favor of a model that depends only on strict ones. From the perspective of the Fisher geometry, this can explicitly be thought of as a ``UV regularization scheme"; a consistent method for dealing with the fact that we cannot resolve up to the arbitrarily small distances in model space necessary to identify between models that differ only along directions coordinatized by sloppy parameters \cite{2011.12420v1, 2111.07176v2}. Thus, one way of interpreting the DB-RG scheme is as an automated data compression algorithm which sequentially integrates out ``high energy" parameters (e.g. parameters associated with small eigenvalues of the Fisher metric) in the same way that a physical RG integrates out large momentum shells. In Section \ref{sec: app} we provide an explicit demonstration of this fact. 

Putting the preceding discussion into more physical language, stiff and sloppy parameters are the data science answer to the idea of relevant and irrelevant operators. In the case where the model of interest \emph{does} possess a ``real" scale, this analogy becomes explicit. For example in the physics literature \cite{Balasubramanian:2014bfa} observed that in the space of conformal field theories, the Fisher metric coincides with the Zamolodchikov metric and thus the hierarchy of relevant and irrelevant operators as dictated by the spectrum of the latter coincides with the hierarchy of stiff and sloppy parameters as dictated by the former. In a similar vein, but working in the opposite direction \cite{PhysRevLett.126.240601} found that the parameters that were regarded as most significant by the Information Bottleneck formalism for data compression \cite{tishby2000information} coincide with the most relevant operators in the conventional RG sense provided the model that is being compressed is given by a local statistical field theory. 

In light of these observations and the information theoretic character of the emergent scale in DB-RG, another central goal of this note is to encourage the reader to think about renormalization as a manifestly information theoretic procedure. For example, consider two theories that differ only in terms of modes that exist above some observable momentum scale (ie the UV cutoff). For all intents and purposes these theories are equivalent, as there is no existing experiment which can be conducted to differentiate them. As we have now established, there is a clear analog to this in broad data science contexts: one has two models that differ only along “sloppy directions” in the model manifold. Sloppy parameters cannot be tuned except with a quantity and/or precision of observation that is not achievable due to experimental limitation. Thus, again, such models should be regarded as practically equivalent. Ultimately, this line of thinking suggests that an RG universality class should correspond to the set of all models/theories which yield equivalent predictions below a threshold set by the amount of useful information that can be collected about the system in question. From this perspective, the relevant notion of scale is always the distinguishability in model/theory space, it just happens that the same information can be communicated in terms of an energy scale in the physical case because such scales bound our experimental capability.\footnote{We should note that similar ideas to this have been communicated before in the physics literature by \cite{Beny:2012qh, Beny:2014sna}, and in the data science literature by \cite{Raju_2018}.} 

The organization of the paper is as follows. In Section \ref{sec: ERG} we review Exact Renormalization in its original physical context. We stress the perspective that a useful subclass of ERG schemes constitute functional diffusion equations, as was originally touted by \cite{cotler2022renormalization}, and recognize the role played by the physical scale in defining such diffusive ERGs. Building on the viewpoint that renormalization and diffusion are equivalent we identify diffusion as a useful device for renormalizing data models even without physical scales in Section \ref{sec: Diffusion}. This picture of diffusion based renormalization for general data models is closely related to the influential diffusion learning paradigm \cite{sohl2015deep} in which intractable distributions are run through diffusion channels in order to produce tractable models for data generation tasks. However, in the absence of a physical scale one cannot control the information which is coarse-grained out of the model, in contrast with physical renormalization schemes which always remove information in a hierarchy of real energy scales. This motivates the Bayesian Renormalization scheme which is introduced in Section \ref{sec: Bayes}. After reviewing the Fisher geometry underpinning Bayesian inference (Section \ref{sec: infogeo}), and the Dynamical Bayesian Inference scheme (Section \ref{sec: dynamicalbayes}), we explicitly derive the DB-RG scheme in Sections \ref{sec: Model RG} and \ref{sec: Data RG}. The DB-RG scheme can be understood as a specific form of diffusive renormalization in which the information that is lost to diffusion is governed by the Fisher metric. Hence, DB-RG is a precisely the refined form of diffusion learning we sought in which the inverse of the distance measure induced by the Fisher metric plays the role of an energy scale for the purpose of the ERG. In Section \ref{sec: app} we illustrate the usefulness of our approach by applying the philosophy of Bayesian Renormalization to a simple autoencoder. We conclude with discussion in Section \ref{sec: Discussion}, in which we review our new perspective on renormalization and suggest future directions both for applications of DB-RG to data science tasks, and as a new theoretical tool.  

\section{Renormalization and Diffusion}

In this section we review exact renormalization in its original physical context with an emphasis placed on the relationship between renormalization and diffusion. We establish that a renormalization group flow corresponds to a semigroup of conditional expectation operators acting on a sample space of random variables appropriate to a given theory. These conditional expectations generate a Markov process that can be associated with a stochastic differential equation, or equivalently with a partial differential equation of the Fokker-Planck form. The stochastic differential equation dictates a coarse-graining scheme in the usual sense, while the partial differential equation absorbs the impact of this coarse graining at the level of the probability distribution describing the variables of interest. 

In a physical theory, a renormalization group flow is seeded with information about the hierarchy of momentum scales through the so-called ERG kernel, which ensures that the information coarse grained away by diffusion is associated with high energy data. By contrast, naive diffusion based renormalization discards information indiscriminately in systems that do not possess a physical scale. This motivates the Bayesian Renormalization scheme discussed in Section \ref{sec: Bayes}, which establishes a meaningful scale for arbitrary systems of random variables that can be seeded into the analog of the ERG kernel. This scale is associated with the distinguishability of probability distributions in the space of models/theories, and reproduces a physical scale whenever one is present.

\subsection{Exact Renormalization is Diffusion} \label{sec: ERG}

Following the lead of \cite{cotler2022renormalization}, we take the perspective that an exact renormalization group flow can be understood as a one parameter family of probability distributions, $\{P_{\Lambda}[\phi]\}_{\Lambda \in \mathbb{R}}$. Here $\Lambda$ is a physically meaningful RG scale (typically associated with a momentum cutoff), and $\phi \in \mathcal{F}$ corresponds to the field configuration relevant to a given theory. $P_{\Lambda}[\phi]$ should therefore be read as the probability density assigned to the field configuration $\phi$ at the scale $\Lambda$. Schematically, we regard
\begin{equation}
	P_{\Lambda}[\phi] \propto e^{-S_{\Lambda}[\phi]},
\end{equation}
where $S_{\Lambda}[\phi]$ is the renormalized action at scale $\Lambda$. The guiding principle of the ERG is that the flow $P_{\Lambda}[\phi]$ must be chosen in such a way that the partition function is preserved:\footnote{Here $\mathcal{D}\phi$ is the path integral measure. For a review of path integral techniques and functional renormalization see \cite{cardy2010introduction}, or for a more thorough treatment, see \cite{peskin1996}.}
\begin{equation} \label{ERG principle}
	\frac{d}{d\ln \Lambda} \int_{\mathcal{F}} \mathcal{D}\phi \; P_{\Lambda}[\phi] = 0.
\end{equation}
The ERG principle \eqref{ERG principle} ensures that all correlation functions below the scale $\Lambda$ are preserved over the course of the ERG flow. 

The most familiar form of exact renormalization is the so-called Polchinski scheme \cite{polchinski1984renormalization}. In Polchinski's ERG, one writes the probability distribution over fields in the form
\begin{equation} \label{Polchinski Flow}
	P_{\Lambda}[\phi] \propto e^{-\frac{1}{2} \int \frac{d^d p}{(2\pi)^d} \; \phi(p) G(p^2) K^{-1}_{\Lambda}(p^2) \phi(-p)} e^{-S_{int,\Lambda}[\phi]}.
\end{equation}
We recognize the first term as the Gaussian distribution associated with a free field theory with propagator $G(p^2)$, but with the incorporation of a function $K^{-1}_{\Lambda}(p^2)$ which plays the role of a smooth cutoff function in momentum space. In words, $K_{\Lambda}^{-1}(p^2)$ suppresses the contribution of momentum modes above the cutoff scale $\Lambda$. The second term in \eqref{Polchinski Flow} is the exponential of the renormalized interacting action at the scale $\Lambda$. 

In Polchinski's picture, $K_{\Lambda}(p^2)$ has a prescribed dependence on $\Lambda$, thus Polchinski's ERG equation arises by determining the equation which must be obeyed by $S_{int,\Lambda}[\phi]$ in order to satisfy the principle \eqref{ERG principle}. By a straightforward computation, one can show that the resulting equation can be put into the form:
\begin{flalign} \label{ERG Fokker Planck Polchinski}
	\frac{d}{d\ln\Lambda} P_{\Lambda}[\phi] &= \int_{M \times M} d^dx d^dy \; \left\{{C}^{Pol.}_{\Lambda}(x,y) \frac{\delta^2 P_{\Lambda}[\phi]}{\delta\phi(x) \delta\phi(y)} + \frac{\delta}{\delta \phi(x)}\left(P_{\Lambda}[\phi]{C}^{Pol.}_{\Lambda}(x,y) \frac{\delta V^{Pol.}_{\Lambda}[\phi]}{\delta \phi(y)}\right)\right\}    \\
	&\equiv \Delta P_{\Lambda}[\phi] + \text{div}\left(P_{\Lambda}[\phi] \text{grad}_{{C}^{Pol.}_{\Lambda}} V^{Pol.}_{\Lambda}[\phi]\right), 
\end{flalign}
where
\begin{equation} \label{Polchinski ERG}
	{C}^{Pol.}_{\Lambda}(p^2) = (2\pi)^d G(p^2)^{-1} \frac{\partial K_{\Lambda}(p^2)}{\partial \ln \Lambda}; \;\;\; V^{Pol.}_{\Lambda}[\phi] = \int \frac{d^dp}{(2\pi)^d} \phi(p) G(p^2) K^{-1}_{\Lambda}(p^2) \phi(-p).
\end{equation}
One might recognize \eqref{ERG Fokker Planck Polchinski} as the \emph{Fokker-Planck} equation with diffusion governed by ${C}^{Pol.}_{\Lambda}(p^2)$ and drift governed by the potential $V^{Pol.}_{\Lambda}[\phi]$. This is the first indication of a deep relationship between exact renormalization and diffusion. Note that the equivalence between (4) and (5) is just a rewriting in terms of the functional (infinite dimensional) equivalent of vector operators. This is so one can identify \eqref{ERG Fokker Planck Polchinski} as a functional version of Fokker-Plank.
 
We refer to the Polchinski approach as an \emph{ERG Scheme} because it corresponds to a particular \emph{choice} on how to renormalize the theory described by \eqref{Polchinski Flow}. The nexus of this choice can be traced back to the way Polchinski decided to regulate the action in \eqref{Polchinski Flow} by introducing a smooth cutoff function $K^{-1}_{\Lambda}(p^2)$. This choice manifests itself in the particular form of the diffusion and drift aspects of \eqref{Polchinski ERG} specifying the Fokker-Planck equation associated with the Polchinski ERG \eqref{ERG Fokker Planck Polchinski}. Choosing different regulating functions corresponds to different ERG schemes, and as a result different Fokker-Planck equations specified by the data \eqref{Polchinski ERG}.

More abstractly, we can \emph{define} an ERG directly by specifying the data $({C}_{\Lambda},V_{\Lambda})$ corresponding to the diffusivity and drift of a Fokker-Planck equation. Equation \eqref{ERG Fokker Planck Polchinski} still holds but with $({C}_{\Lambda},V_{\Lambda})$ replacing ${C}^{Pol.}_{\Lambda}(p^2), V^{Pol.}_{\Lambda}[\phi]$. This is then an ERG corresponding to a different scheme.

The Fokker Planck equation corresponds to a bonafide ERG because it satisfies the ERG principle \eqref{ERG principle}. To see that this is the case, let us now show that we can rewrite \eqref{ERG Fokker Planck Polchinski} in the form
\begin{equation} \label{WM Equation}
	-\frac{d}{d\ln\Lambda} P_{\Lambda}[\phi] = \int_M d^dx \; \frac{\delta}{\delta \phi(x)} \left(\Psi_{\Lambda}[\phi;x] P_{\Lambda}[\phi]\right),
\end{equation} 
where $M$ is the spacetime manifold on which the theory is defined \cite{morris1994exact}. Hopefully it is clear that any one parameter family $P_{\Lambda}[\phi]$ satisfying \eqref{WM Equation} also satisfies \eqref{ERG principle}. This is because \eqref{WM Equation} specifies a \emph{divergence flow}, that is the right hand side of \eqref{WM Equation} is a divergence in the space of field configurations. We can therefore employ the divergence theorem to observe that
\begin{equation} \label{WM is RG}
	\frac{d}{d\ln\Lambda} \int_{\mathcal{F}} \mathcal{D} \phi \; P_{\Lambda}[\phi] = - \int_{\mathcal{F}} \mathcal{D}\phi \; \int_M d^dx \; \frac{\delta}{\delta \phi(x)} \left(\Psi_{\Lambda}[\phi;x] P_{\Lambda}[\phi]\right) = 0.
\end{equation}

In order to write \eqref{ERG Fokker Planck Polchinski} in the form \eqref{WM Equation} we take
\begin{equation} \label{Reparameterization Kernel}
	\Psi_{\Lambda}[\phi;x] = \int_M d^dy \; {C}_{\Lambda}(x,y) \frac{\delta \Sigma_{\Lambda}[\phi;P_{\Lambda}]}{\delta \phi(y)},
\end{equation}
as has appeared previously in \cite{rosten2012fundamentals, morris1994exact, Morris:2006in, cotler2022renormalization, Matsumoto_2020}. Here $C_{\Lambda}(x,y)$ is the \emph{ERG kernel} appearing in the Fokker-Planck equation associated to the ERG, and $\Sigma_{\Lambda}[\phi;P_{\Lambda}]$ is called the \emph{scheme functional} which is determined through the \emph{ERG potential} $V_{\Lambda}$ via the equation
\begin{equation} \label{ERG Potential}
	\Sigma_{\Lambda}[\phi;P_{\Lambda}] = -\ln\left(\frac{P_{\Lambda}[\phi]}{e^{-V_{\Lambda}[\phi]}}\right) = S_{\Lambda}[\phi] - V_{\Lambda}[\phi].
\end{equation}
Plugging \eqref{Reparameterization Kernel} back into \eqref{WM Equation}, we reconcile \eqref{ERG Fokker Planck Polchinski} with the diffusion and drift aspects given by $(C_{\Lambda}, V_{\Lambda})$, as desired. Together $(C_{\Lambda},V_{\Lambda})$ therefore specify a consistent scheme for regulating the high energy degrees of freedom of the field theory, in analogy with the regulating function $K^{-1}_{\Lambda}(p^2)$ appearing in \eqref{Polchinski Flow}. 

It is worth noting that \eqref{WM Equation} defines an approach to ERG that is more general than diffusion. All divergence flows, which can generically be written in the form \eqref{WM Equation}, specify ERG flows as evidenced by \eqref{WM is RG}. However, only the subset of divergence flows in which the \emph{reparamterization kernel} $\Psi_{\Lambda}$ is taken to be of the form \eqref{Reparameterization Kernel} result in Fokker-Planck equations. Equation \eqref{WM Equation} is called the \emph{Wegner-Morris equation}, and ERG schemes satisfying the Wegner-Morris equation are called \emph{Wegner-Morris schemes}. The choice of Wegner-Morris scheme is encapsulated entirely in the reparameterization kernel. We shall refer to reparameterization kernels which are in the form of \eqref{Reparameterization Kernel} as \emph{Fokker-Planck schemes} to highlight their relationship with diffusion. A Fokker-Planck scheme is specified entirely by the data $(C_{\Lambda},V_{\Lambda})$. 

To conclude this section, let us briefly explore an alternative way to recognize that the Wegner-Morris equation specifies an ERG flow which allows us to supply a more concise and intuitive interpretation of the ERG. The effect of (\ref{WM Equation}) on $P_{\Lambda}$ can be absorbed by continuously reparameterizing the fields $\phi$ at each new scale according to the rule
\begin{equation} \label{Reparameterization}
	\phi'(x) = \phi(x) + (\delta \ln\Lambda) \Psi[\phi;x].
\end{equation}
Equation (\ref{Reparameterization}) should be regarded as the integral curve of $\Psi_{\Lambda}[\phi;x]$ in the space of field configurations, where we are regarding $\Psi_{\Lambda}[\phi;x] \in T\mathcal{F}$ as a tangent vector to this space. Solving (\ref{Reparameterization}) results in a one parameter family of field configurations $\{\phi_{\Lambda}\}_{\Lambda \in \mathbb{R}}$, in which $\phi_{\Lambda}$ can be thought of as describing the relevant field degrees of freedom at scale $\Lambda$. In this way, exact renormalization can be connected with more familiar Wilsonian renormalization schemes by interpreting equation \eqref{Reparameterization} as specifying a coarse graining procedure. The coarse graining map (\ref{Reparameterization}) is a diffeomorphism in the space of field configurations, which means it must leave the integral
\begin{equation}
	\int_{\mathcal{F}} \mathcal{D}\phi \; P_{\Lambda}[\phi]
\end{equation}
invariant. Thus, again, we find that the equation \eqref{WM Equation} specifies a meaningful renormalization scheme in the usual sense of satisfying \eqref{ERG principle}.

Specializing to Fokker-Planck ERG schemes, we can expand on this discussion. As was introduced in detail in \cite{berman2022inverse}, a (functional) Fokker-Planck equation of the form (\ref{ERG Fokker Planck Polchinski}) is associated with a (functional) stochastic differential equation:
\begin{equation} \label{ERG Stochastic Differential equation}
	d\phi(x) = -\text{grad}_{{C}_{\Lambda}} V_{\Lambda}[\phi] (d\ln\Lambda) + \sqrt{2} \int_{M} d^dy \; \sigma_{\Lambda}(x,y) dW_{\Lambda}(y).
\end{equation}
Here, $W_{\Lambda}(x)$ is a function valued Weiner process, and $\sigma_{\Lambda}$ is the diffusivity kernel defined by the property that it ``squares" to the covariance ${C}_{\Lambda}$:
\begin{equation}
	\int_{M} d^dz \; \sigma_{\Lambda}(x,z) \sigma_{\Lambda}(z,y) = {C}_{\Lambda}(x,y).
\end{equation}
Equation (\ref{ERG Stochastic Differential equation}) is the stochastic differential equation that arises from the deterministic gradient flow defined by (\ref{Reparameterization}) subject to noise with covariance governed by ${C}_{\Lambda}$. Thus, we have arrived at the punchline: An exact renormalization group flow specified by the data $(\mathcal{F},{C}_{\Lambda}(x,y),V_{\Lambda}[\phi])$ can be understood as the stochastic coarse graining of the field degrees of freedom arising from the reparameterization of the fields under the gradient of the potential $V_{\Lambda}$ subject to noise governed by ${C}_{\Lambda}$.

\subsection{Diffusion is Exact Renormalization} \label{sec: Diffusion}

In Section \ref{sec: ERG}, we illustrated how exact renormalization group flows a la Wegner-Morris correspond to a subclass of functional diffusion equations \eqref{WM Equation}. These Fokker-Planck schemes are specified by the pair $({C}_{\Lambda}, V_{\Lambda})$ which set, respectively, the diffusion and drift of a stochastic process underyling the Fokker-Planck equation associated with the ERG. The stochastic process \eqref{ERG Stochastic Differential equation} explicitly describes a coarse graining of the field degrees of freedom, while the related diffusion equation encodes the impact of such a coarse graining on the renormalized action, resulting in an effective field theory at each scale. One of the main insights of \cite{berman2022inverse} is that viewing exact renormalization as a diffusion process suggests an approach to renormalization which is applicable to a wider class of systems modeled by probability distributions, beyond merely those which posses a physical interpretation. In this section we will outline this approach.

Consider a random variable $Y$ with sample space $S$. For simplicity, let us assume that $Y$ is a continuous random variable, and $S$ is a Riemannian manifold. A general class of diffusion equations on $S$ are then specified by a one parameter family of probability distributions $\{p_t(y)\}_{t \in \mathbb{R}}$ along with a metric\footnote{Here, $TS$ is the tangent bundle to the manifold $S$.} $g_t: TS \times TS \rightarrow \mathbb{R}$ and a potential function $V_t: S \rightarrow \mathbb{R}$ such that
\begin{equation} \label{FD Fokker-Planck}
	\frac{dp_t(y)}{dt} = \Delta p_t(y) + \text{div}\left(p_t(y) \text{grad}_{g_t} V_t(y)\right). 
\end{equation}
Equation (\ref{FD Fokker-Planck}) is the Fokker-Planck equation associated with the stochastic differential equation
\begin{equation} \label{FD Stochastic DiffEQ}
	dY^i_t = -\left(\text{grad}_{g_t} V_t \right)^i dt + \sqrt{2} (\theta_t)^i_j dW^j_t
\end{equation}
where\footnote{Those who are familiar may notice that Eqn. \eqref{veilbeins} identify $(\theta_t)^i_k$ (at each $t$) as the components of a vielbeins for the metric $g_t$.}
\begin{equation} \label{veilbeins}
\delta^{kl} (\theta_t)^i_k (\theta_t)^j_l = g_t^{ij}.
\end{equation}
It is worth noting here that the $g_t$ plays a double role. From one perspective it is the metric in this process but interpreted in terms of the underlying statistics it is the inverse of the covariance matrix.
 Then equations \eqref{FD Fokker-Planck} and \eqref{FD Stochastic DiffEQ} should be compared with \eqref{ERG Fokker Planck Polchinski} and \eqref{ERG Stochastic Differential equation} from the exact renormalization context. 

As was the case in Section \ref{sec: ERG}, we can formally solve for the gradient flow of the potential $V_t$ to determine a one parameter family of renormalized degrees of freedom. To be precise, let $\gamma: \mathbb{R} \rightarrow S$ be a one parameter family of points in $S$ solving the gradient flow problem
\begin{equation}
	\frac{d\gamma_t}{dt} = \text{grad}_{g_t} V\rvert_{\gamma_{\tau}}.
\end{equation}
In terms of $\gamma_t$ we can now write \eqref{FD Stochastic DiffEQ} in the form
\begin{equation}
	dY^i_t = -\dot{\gamma_t}^i dt + \sqrt{2} (\theta_t)^i_j dW^j_t.
\end{equation}
where $\dot{\gamma}_t$ and $\theta_t$ correspond to the drift and diffusion, respectively. 

The Fokker-Planck equation (\ref{FD Fokker-Planck}) has a schematic solution:
\begin{equation} \label{Diffusion Process}
	p_t(y) = \int_S d^d y_0 \; \pi(y,y_0;t) p_0(y_0).
\end{equation}
Here, $p_0(y)$ is the initial data, and $\pi(y,y_0;t)$ is the \emph{Heat Kernel}. Provided the drift and diffusion are constant over the full sample space the diffusion kernel will be a  Gaussian of the form\footnote{Strictly speaking this form of the heat kernel is only approximate to leading order in the rate of change of $g_{t}$, however we can always control how fast $g_{t}$ changes in order to make sure this form holds to arbitrary precision.}
\begin{equation}
	\pi(y,y_0;t) = \mathcal{N}(\gamma_t, tg_t^{-1})(y-y_0)
\end{equation}
Explicitly, $\pi(y,y_0;t)$ is the transition probability density for a sample point to start from $y_0$ and diffuse to $y$ in a time $t$. We shall interpret the heat kernel as a stochastic map encoding the implicit coarse graining scheme associated with the diffusive RG flow. 

This picture of renormalization is closely related to information theoretic approaches to renormalization from the high energy physics community that have been employed in the study of holography and operator algebras \cite{faulkner2020holographic,furuya2020real,Gesteau:2023hbq}. To see this, note that we can encode \eqref{Diffusion Process} as a semigroup of \emph{conditional expectation operators} acting on the space of functions on $S$, $E_{t}: \Omega^0(S) \rightarrow \Omega^0(S)$. The conditional expectation operator $E_t$ acts as\footnote{In Eqn. \eqref{Exp conv} and hereafter we shall use the following standard notation: When taking the expectation value of a function of a random variable, $X \sim p$, the random variable shall appear as a capital letter inside of the expectation and the probability distribution shall appear as a subscript e.g., $\mathbb{E}_p\big(f(X)\big)$.}
\begin{equation} \label{Exp conv}
	E_{t}(f) = \mathbb{E}_{\pi_{t}}(f(Y_0)) = \int_{S} \text{Vol}_S(y_0) \; \pi(y,y_0;t)f(y_0),
\end{equation}
so that, for example, the posterior predictive distribution is given by
\begin{equation}
	p_t(y) = E_t(p_0)(y) = \mathbb{E}_{\pi_t}(p_0(Y_0)).
\end{equation}
The set of operators $\{E_t\}_{t \in \mathbb{R}}$ form a semigroup in the sense that
\begin{equation}
	E_{t_2} \circ E_{t_1} = E_{t_1 + t_2}.
\end{equation}

In a more general context, a conditional expectation on a von Neumann algebra $M$ is a projection of $M$ to a subset $N \subset M$, $E: M \rightarrow N$, which retains the normalization of states such that $E(\mathbb{1}_M) = \mathbb{1}_N$ \cite{takesaki1972conditional}. A very broad class of renormalization schemes accessible to quantum probabilities can subsequently be formulated as a semigroup of conditional expectation operators $\{E_{\Lambda}\}_{\Lambda \in \mathbb{R}}$ acting on the space of operators affiliated with a given system. In a recent work \cite{Gesteau:2023hbq}, this form of renormalization was given the name \emph{code subspace renormalization} to reflects its relationship with error correction.\footnote{In the case that the operator algebra in question is Abelian, the formal definition of a conditional expectation is in one to one correspondence with the set of conditional probability distributions in the ordinary measure theoretic sense. This suggests that one can obtain an explicit form for the renormalization of quantum states (viewed as states on non-Abelian operator algebras) in terms of generalized diffusion processes as discussed in \cite{carlen2014analog,carlen2017gradient,carlen2020non}. We plan to explore these generalizations in forthcoming work.}

The relationship between renormalization and diffusion is very satisfying because we can think of a diffusion process as destroying some of the fine grained information stored in a very complicated, ``UV complete" probability distribution. Indeed, this is the point of view which is advocated for in the influential diffusion learning paradigm \cite{sohl2015deep}. In diffusion learning, one begins with a highly complex and intractable distribution $p_0$ which is run through a forward diffusion process for a time $t_f$ in order to arrive at an analytic distribution, $p_f$. One can then sample data from $p_0$ by first sampling data from $p_f$ and subsequently solving a ``reverse" stochastic differential equation derived from \eqref{FD Stochastic DiffEQ} \cite{song2020score}. To be more precise, given a generic forward diffusion process specified by the SDE
\begin{equation} \label{Forward Diffusion}
	dY_t^i = \mu^i_t(Y_t) dt + \sigma_t{}^i{}_j(Y_t) dW_t^j,
\end{equation}
the associated reverse SDE is of the form \cite{anderson1982reverse}
\begin{equation} \label{Reverse Diffusion}
	dY_{s}^i = \bigg\{ \mu^i_{s} - \frac{1}{2} \partial_j \bigg(\delta^{kl} \sigma_{s}{}^i{}_k \sigma_{s}{}^j{}_l\bigg) - \delta^{kl} \sigma_{s}{}^i{}_k \sigma_{s}{}^j{}_l \partial_j \ln p_{s} \bigg\}\bigg\rvert_{Y_{s}} ds + \sigma_s{}^i{}_j(Y_{s}) dW_{s}^j
\end{equation}
Here, $t$, which runs from $0$ to $t_f$, is the forward time, while $s$, which runs from $t_f$ to the initial time $0$, is the reverse time. The reverse SDE is determined by all of the same data as the forward SDE with exception of the score functions -- $\partial_j \ln p_{s}(y)$. Here, $p_{s}(y)$ is probability distribution obtained by diffusing $p_0$ according to  the forward SDE up to the reverse time $s$. Simulating the reverse SDE therefore amounts to efficiently reconstructing the scores. In the machine learning community, a resolution to this problem has been presented in terms of the training of a score based generative algorithm (see \cite{song2020score} for a comprehensive review). Interestingly, such a score-based model can be interpreted as a statistical inference problem in which the distribution $p_{s}$ is \emph{learned} by observing draws from intermediate diffused distributions. We will revisit this in Section \ref{sec: Bayes} where we provide an alternative point of view on the same fact by demonstrating that inverting a diffusion process can be thought of as a Bayesian inference experiment.

Based on the observations of the preceding paragraph, one may be tempted to say that diffusion learning can be thought of as a form of exact renormalization for data science models. In light of the correspondence between ERG and diffusion, such a statement is technically correct. However, naive diffusion lacks a very crucial feature which is at the core of a good renormalization group flow; namely a meaningful notion of scale. Indeed, unlike physical renormalization which coarse grains hierarchically over length/energy scales, naive diffusion removes information in an essentially unstructured way. Thus, while basic diffusion can technically be regarded as a form of renormalization, it fails to provide real control over how a probability model changes as a function of any meaningful scale. More plainly, in physics based renormalization we know that the information lost to coarse graining correspond to high energy modes via the construction of the ERG kernel. By contrast, naive diffusion of a probability distribution apparently discards information indiscriminately. Ultimately, this is problem that the Bayesian approach to diffusion/renormalization will overcome.

\section{Bayesian Renormalization and Information Geometry} \label{sec: Bayes}

We now turn to an information theoretic approach to renormalization which will provide a meaningful scale that is applicable to arbitrary random variables, and reproduces the physical scale when the chosen random variable descends from a physical system. This notion of scale is defined in terms of the Fisher information metric, and therefore corresponds to the distinguishability between points in the space of probability distributions. 
Our new approach emphasizes the role of information and information geometry in renormalization, which we now understand broadly as a mechanism for identifying equivalence classes of probability distributions that are indistinguishable as predictive models at a level of precision fixed by the amount of accessible data. This viewpoint allows us to draw a very sharp analogy between renormalization and aspects of data compression \cite{legeza2004quantum,Kingma:2013hel,Machta_2013,kingma2021variational, PhysRevLett.126.240601}, data generation \cite{gui2021review,ramesh2021zero, ramesh2022hierarchical}, data classification \cite{2301.02683v1, Bansal2022, 
fischer2022decomposing, Fleig_2022, Rodriguez_Nieva_2019, neal2001annealed, Raju_2018}, dimensional reduction and model selection \cite{2102.06701v1,2011.12420v1,2111.07176v2} commonly studied in data science and machine learning. We present this approach to renormalization through its relationship with Bayesian inference to highlight its information theoretic origin. 

Bayesian inference is an approach to reconstructing the probability distribution responsible for generating a sequence of observed data. Let $Y$ be a random variable taking values in the sample space $S$. Given a series of independent draws, $\{y_t\}_{t = 1}^T$, from the data generating distribution $p^*_Y$, the output of a Bayesian inference is a posterior predictive distribution, $p_T(y)$, which is the best approximation to $p^*_Y$ given the data that has been observed. The idea of Bayesian renormalization is to perform Bayesian inference \emph{in reverse} by discarding data rather than incorporating it. The posterior predictive distribution subsequently pools attributes from models that are similar but not equivalent to the data generating model. As we shall illustrate, this defines an RG scheme that automatically encodes an information theoretic notion of relevant and irrelevant degrees of freedom. It is specified by an explicit diffusion process that enforces the relevance criterion by coarse-graining in a way that sequentially ``integrates out" parameters in order of their relevance just like one would sequentially integrate over momentum shells in a Wilsonian renormalization.

\subsection{Bayesian Inference and Information Geometry} \label{sec: infogeo}

The first step in Bayesian inference is \emph{model selection}. This corresponds to choosing a parametric family of probability distributions on $S$, $p_{Y \mid \Theta}(y \mid \theta)$. A priori, we should consider any allowed distribution for $Y$ as a candidate for the data generating distribution. Thus, we might take
\begin{equation}
	p_{Y \mid \Theta}(y \mid \theta) = e^{-\theta^i S_i(y)}
\end{equation}
where here $\{S_j(y)\}_{j \in \mathcal{J}}$ is the set of log-likelihoods corresponding to allowed probability models for $Y$. In principle the space of models might be infinite dimensional, however for simplicity let us assume that it is finite.\footnote{The following analysis carries over formally to the infinite dimensional case.} 

Let $\mathcal{M} = \{p_{Y\mid \Theta}(y \mid \theta) \; | \; \theta \in \mathbb{R}^n\}$. In words, $\mathcal{M}$ is a space whose points correspond to probability distributions for $Y$. By construction, this space has a local coordinate system given in terms of the parameters $\theta$. A natural basis for the tangent space of $\mathcal{M}$, is given to us in terms of the \emph{score vectors} $\underline{\ell}_i = \frac{\partial}{\partial \theta^i} \ln(p_{Y\mid\Theta}(y\mid\theta))$. It is easy to see that these vectors are linearly independent and spanning insofar as they are isomorphic to the coordinate basis $\underline{\partial}_i$. Moreover, when viewed as functions on $S$ the score vectors have zero expectation value due to the normalization condition on $p$:
\begin{equation}
	\mathbb{E}_\theta(\underline{\ell}_i(Y)) = \int_S d^dy \; p_{Y\mid\Theta}(y \mid \theta) \frac{\partial \ln(p_{Y\mid\Theta}(y\mid\theta))}{\partial \theta^i} = \frac{\partial}{\partial \theta^i} \int_S d^dy \; p_{Y\mid\Theta}(y\mid\theta) = 0.
\end{equation}
In terms of this basis, we define the Fisher information matrix:
\begin{equation} \label{Fisher Matrix}
	\mathcal{I}_{ij}(\theta) = \mathbb{E}_{\theta}(\un{\ell}_i \un{\ell}_j) = \mathbb{E}_{\theta}\left(\frac{\partial \ln(p_{Y\mid\Theta}(Y\mid\theta))}{\partial \theta^i} \frac{\partial \ln(p_{Y\mid\Theta}(Y\mid\theta))}{\partial \theta^j}\right).
\end{equation}
The Fisher matrix \eqref{Fisher Matrix} should be interpreted as the components of a metric, the Fisher metric, on $\mathcal{M}$ in the basis $\{\un{\ell}_i\}$.

The Fisher metric provides an infinitesimal measure of the similarity between two models in $\mathcal{M}$. This can be seen most clearly through the relationship between the Fisher metric and the KL-divergence. Recall,
\begin{equation}
	D_{KL}(\theta \parallel \theta') = \mathbb{E}_{\theta}\left(\ln(\frac{p_{Y\mid\Theta}(Y\mid\theta)}{p_{Y\mid\Theta}(Y\mid\theta')}) \right).
\end{equation}
measures the relative entropy between two distributions. $D_{KL}$ is an information divergence, which means that $D_{KL}(\theta \parallel \theta') \geq 0$ and it is equal to zero if and only if $\theta = \theta'$. This makes $D_{KL}$ a good measure of the distinguishability between models. In general, however, $D_{KL}$ is not a symmetric function of $\theta$ and $\theta'$ and therefore cannot be regarded itself as an inner product. Nonetheless, in the immediate neighborhood of a point $\theta \in \mathcal{M}$ the KL-divergence can be expanded to quadratic order as
\begin{equation}
	D_{KL}(\theta \parallel \theta') = \frac{1}{2} \mathcal{I}_{ij}(\theta) \delta \theta^i \delta \theta^j + \mathcal{O}(\delta \theta^3). 
\end{equation}

The Fisher metric plays a significant role in parameter estimation because it encodes the sensitivity of a model's output to changes in parameter values \cite{Raju_2018}. This is closely related to the task of distinguishing between model parameters that are ``sloppy" and ``stiff" \cite{daniels2008sloppiness,transtrum2015perspective}. Strict parameters covary strongly with the model output and therefore correspond to large eigenvalues of the Fisher metric. Sloppy parameters, on the other hand, covary only weakly with the model output and therefore correspond to small eigenvalues of the Fisher metric. From a geometric perspective, this means that we can think of sloppy directions -- that is directions coordinatized by sloppy parameters -- as being highly compact in the space of models. In particular, two models that differ only along sloppy directions will be very hard to distinguish. As a consequence, sloppy parameters require an extensive amount or quality of observed data in order to be fit \cite{abbott2022far}. However, because these parameters only weakly impact the predictive power of the model it may also be possible to systematically remove them in favor of a reduced model that depends only a sufficient set of strict parameters. 

This state of affairs strongly resembles the typical use case of the renormalization group in physical theories. Our ability to meaningfully distinguish between theories is capped by our capacity to perform experiments which measure physics beyond particular scales. In this sense, two theories which yield equivalent predictions except beyond scales that cannot be experimentally probed must be regarded as practically equivalent. If we view a physical theory as parameterized by the Fourier modes of field degrees of freedom, it is precisely the high energy modes which correspond to the ``sloppy parameters" since one requires an extensive or even impossible set of measurements to distinguish between theories that differ only at very high energy scales. From this point of view, we recognize renormalization as a scheme for systematically regulating sloppy model parameters in order to arrive at an equivalence class of models that cannot be distinguished at the level of accuracy admitted by our present ability to observe data. 

In recent work \cite{2111.07176v2} it has been suggested that a similar approach would be very useful in a more broad data science context. Quite serendipitously, this can be thought of as a ``UV regularization scheme" from the perspective of the Fisher geometry -- it is a consistent method for dealing with the fact that we cannot resolve up to the arbitrarily small distances in model space necessary to identify between models that differ along directions coordinatized by sloppy parameters. As we introduce the Bayesian renormalization scheme, we will highlight how it \emph{automatically} performs an information geometrically natural regularization to this end, that is oriented directly towards intelligently removing these sloppy degrees of freedom.

\subsection{Dynamical Bayesian Inference} \label{sec: dynamicalbayes}

The upshot of Section \ref{sec: infogeo} is that the starting point of a Bayesian inference experiment can be understood as the specification of a Riemannian information geometry $(\mathcal{M},\mathcal{I})$, where $\mathcal{M}$ consists of all possible probability models for $Y$, and $\mathcal{I}$ is an infinitesimal measure of the distinguishability between models. As we have discussed, adopting an information geometric approach to Bayesian inference already allows us to begin formulating the correspondence between learning in the space of models and renormalization. The next step in Bayesian inference is updating, which we regard as a specification of dynamics in the space of models.

In conventional Bayesian inference, the updating phase starts by specifying a \emph{prior} distribution $\pi_0: \mathcal{M} \rightarrow \mathbb{R}$ which acts as the initial data for the dynamical system. In light of observed data, the prior distribution is updated to the posterior distribution, $\pi_{T}(\theta)$, by using Bayes' law. That is
\begin{equation} \label{Bayesian update}
	\pi_T(\theta) \propto \left(\prod_{t = 1}^T p_{Y\mid\Theta}(y_t \mid \theta)\right) \pi_0(\theta).
\end{equation} 
The constant of proportionality can be deduced by enforcing that $\pi_T$ be a normalized probability distribution on $\mathcal{M}$. One should interpret (\ref{Bayesian update}) as specifying that the probability the data generating model lives in a small neighborhood of the point $\theta \in \mathcal{M}$ is proportional to the probability that one would have observed the existing sample conditional on the underlying parameter value being in such a neighborhood, multiplied by the prior weight given to that neighborhood. In this way the posterior distribution is slowly trained around the region in model space where the data generating distribution lives. 

In \cite{berman2022dynamical}, we asked the question of what Bayesian inference would look like as a continuous time dynamical system. In other words, we think of data as being continuously observed so that the sample $\{y_t\}_{t = 1}^T$ is growing as a function of the ``time" parameter $T$. We subsequently showed that the posterior distribution is governed by an integro-differential equation
\begin{equation} \label{Dynamical Bayes}
	\frac{\partial \pi_T(\theta)}{\partial T} = -\left(D_{KL}(\theta^* \parallel \theta) - \mathbb{E}_{\pi_T}\left(D_{KL}(\theta_* \parallel \Theta)\right) \right) \pi_T(\theta).
\end{equation}
Here $\theta_*$ is the parameter corresponding to the location of the data generating distribution in $\mathcal{M}$, and we have assumed that this value is unique.\footnote{This assumption ensures that the posterior predictive distribution will converge to the data generating model almost surely.} The equation (\ref{Dynamical Bayes}) has a schematic solution
\begin{equation}
	\pi_T(\theta) \propto e^{-TD_{KL}(\theta_* \parallel \theta)}
\end{equation}
which we interpret as a Bolzmann distribution with ``energy" given by the KL-divergence between the data generating model and a model at $\theta \in \mathcal{M}$. This observation is intimately related with the idea of treating the distinguishability of models as a kind of ``energy" scale. 

At sufficiently late $T$ the posterior distribution will be of the form
\begin{equation} \label{Late T posterior}
	\pi_T(\theta) = \mathcal{N}(\mu_T, \frac{1}{T} \mathcal{I}(\mu_T)^{-1})(\theta).
\end{equation}
Here $\mu_T$ is the $T$-path of the maximum a posterior (MAP) estimate. We regard the specification of $\mu_T$ as defining a new dynamical element. Once the full time path of the MAP has been observed, one can construct a potential function $V: \mathcal{M} \rightarrow \mathbb{R}$ for which $\mu_T$ is \emph{defined} to be a gradient flow
\begin{equation} \label{Bayesian Potential}
	\frac{d}{dT} \mu_T = \text{grad}_{\mathcal{I}} V \rvert_{\mu_T}.
\end{equation}
One can consider $V$ as encoding the details of the sequence with which data was observed in relation to the sequential evolution of the answer to the question of which single model best approximates the data generating distribution. Eventually, as $T \rightarrow \infty$ we expect that $\mu_T \rightarrow \theta_*$. 

One may interpret (\ref{Late T posterior}) as specifying that the posterior distribution $\pi_T$ is localized around the MAP, $\mu_T$, with a characteristic width given by $\frac{1}{T} \mathcal{I}(\mu_T)^{-1}$. Notice, the width of this distribution is shrinking as a function of $T$. In other words, as more and more data is observed, the posterior trains around a smaller and smaller neighborhood of the MAP. This means that the posterior predictive model at time $T$ will become more and more precisely attuned to the specific characteristics of the data generating distribution. 

We conclude that a dynamical Bayesian inference scheme is specified by the triple $(\mathcal{M}, \mathcal{I}, V)$, where $\mathcal{M}$ specified the set of allowed models, $\mathcal{I}$ endows this set with a notion of \emph{scale} in terms of the distinguishability of nearby models, and $V$ encodes the sequence with which data is observed and the trajectory of the MAP as it converges towards the data generating model. This set of data is cosmetically very similar to the data defining an exact renormalization group flow, $(\mathcal{F},{C}_{\Lambda}, V_{\Lambda})$. We will now demonstrate how one can \emph{define} an information theoretic exact renormalization group flow in terms of the dynamical Bayesian inference scheme.

\subsection{Backward Inference and Model Space Renormalization} \label{sec: Model RG}

The \emph{posterior predictive distribution} is obtained by taking the convolution of the posterior and the likelihood model:
\begin{equation} \label{Posterior Predictive}
	p_T(y) = \mathbb{E}_{\pi_T}\left(p_{Y\mid\Theta}(y \mid \Theta)\right) = \int_{\mathcal{M}} \text{Vol}_{\mathcal{M}}(\theta) \pi_T(\theta) p_{Y\mid\Theta}(y\mid\theta). 
\end{equation}
Taking the posterior to be of the form (\ref{Late T posterior}), we can see that (\ref{Posterior Predictive}) is a weighted sum of probability models for $Y$ in which the set of models outside a small neighborhood of the MAP are heavily suppressed. Notice that the posterior distribution is playing a very similar role to the cutoff function in (\ref{Polchinski Flow}), only in the opposite direction. The smooth cutoff suppresses high momentum modes or, in other words, information at short scales. Conversely, as $T$ increases, the posterior distribution suppresses information at large distances as measured by the information geometry on $\mathcal{M}$. This makes sense, as we have dictated renormalization is a form of diffusion in which information is removed from the probability model. By contrast, Bayesian inference incorporates new information with the observation of each new data point. This discrepancy motivates the idea of considering Bayesian inference \emph{in reverse} in which, rather than incorporating new data, an experimenter sequentially removes data from the reconstructed model. We refer to this process as \emph{backward inference}. 

Formally, backward inference corresponds to flowing along the inverse ``time" parameter $\tau = \frac{1}{T}$. In terms of this parameter, 
\begin{equation} \label{Backward Posterior}
	\pi_{\tau}(\theta) = \mathcal{N}(\mu_{\tau}, \tau \mathcal{I}(\mu_{\tau})^{-1})(\theta).
\end{equation}
As $\tau$ increases the width of the posterior distribution \emph{increases} and a larger set of models are meaningfully incorporated into the posterior predictive distribution.\footnote{Strictly speaking (\ref{Backward Posterior}) corresponds to the Bayesian posterior only for sufficiently small $\tau$. However, recall that the prior distribution is irrelevant in a Bayesian update. Thus, we can extend (\ref{Backward Posterior}) to all values of $\tau$, and interpret the distribution it flows to as $\tau \rightarrow \infty$ as some prior distribution that converges to the data generating distribution with sufficient observations. This is also preferable for interpreting (\ref{Backward Posterior}) as a renormalization group flow, since we are interested in what ``low energy" effective models we can get to at late $\tau$.} Over time the set of models which live in a small neighborhood of the data generating model will receive less and less weight in the posterior predictive distribution, as the reconstructed model becomes a weighted sum of a more diverse set of models. Inputting \eqref{Backward Posterior} into \eqref{Posterior Predictive}, we find the $\tau$-dependent predictive distribution:
\begin{equation} \label{RG in model space}
	p_{\tau}(y) = \int_{\mathcal{M}} \text{Vol}_{\mathcal{M}}(\theta) \; \mathcal{N}(\mu_{\tau}, \tau \mathcal{I}(\mu_{\tau})^{-1})(\theta) p_{Y \mid \Theta}(y \mid \theta), 
\end{equation} 
which defines a semigroup of conditional expectation values acting in the space of models. Hence we regard \eqref{RG in model space} as defining a renormalization group flow directly \emph{in the space of models}. 

\subsection{Bayesian Inversion and Data Space Renormalization} \label{sec: Data RG}

Operationally, \eqref{RG in model space} defines a perfectly reasonable renormalization scheme which coarse grains in model space. However, there is utility to translating \eqref{RG in model space} so that it can also be understood as explicitly coarse graining in the space of data realizations, as is more standard in typical renormalization schemes. To accomplish this task, we will restrict our attention to spaces of models that are realized in the context of Bayesian inversion \cite{dashti2017bayesian}. 

The goal of a Bayesian inversion problem is to deduce the signal, $\theta$, that predicated a measured output or data, $y$. The data and signal are related by a map, $y = G(\theta)$, which we regard as a deterministic model. In practice, either due to explicit stocasticity or limitations to measurement precision, we must regard the realized output as a random variable which depends conditionally on the signal as
\begin{equation} \label{Signal process}
	Y \mid \Theta = G(\Theta) + N
\end{equation}  
where $N$ is a random variable that is conditionally independent of $Y$ and encodes the aforementioned noise. One may therefore read \eqref{Signal process} as dictating that $Y$ and $\Theta$ are related by a ``law" $Y = G(\Theta)$ but subject to some random fluctuations governed by $N$. Provided that the noise is distributed with some density $p_N(n)$, we can form the conditional density of $Y$ given $\Theta$ by pulling this measure back by \eqref{Signal process} to obtain
\begin{equation} \label{Bayesian inversion family}
	p_{Y \mid \Theta}(y \mid \theta) = p_N(y - G(\theta)).
\end{equation}

The procedure described above is familiar from conventional approaches to Bayesian inference for modeling complex systems and constructing neural networks. For example, in the most simple case of a feed-forward neural network, the deterministic function $G(\theta)$ is given by $f(x;W,b)$ where $f$ is the neural network architecture specified by its set of weights, $W$, and biases, $b$. In the simplest case of $L^2$ loss, we take
\begin{equation}
	p_{Y \mid \Theta}(y \mid \theta) = \mathcal{N}(0,\sigma^2)(y-G(\theta))
\end{equation}
where $\sigma^2$ is a hyperparameter that sets a scale for the tolerated prediction error. 

Given a parametric family of probability distributions for $Y$ which can be written in the form \eqref{Bayesian inversion family}, we are motivated to rewrite the posterior predictive distribution \eqref{RG in model space} as an integral over $S$ by forming the pushforward measure via the mapping $G$.\footnote{The pushforward operation is well defined at the level of measures even though the map $G$ may not be invertible since we only need for the inverse image of $G$ to be measurable.}
\begin{equation} \label{Post Pred = Diffusion}
	p_{\tau}(y) = \int_{S} \text{Vol}_S(y_0) \; \mathcal{N}(\gamma_{\tau}, \tau K_{\tau}^{-1})(y-y_0) \; p_N(y_0).
\end{equation}
Here $\gamma_{\tau} = G(\mu_{\tau})$ is the data prediction associated with the maximum a posteriori estimate, and 
\begin{equation}
	(K_{\tau}^{-1})^{ab} = \frac{\partial G^a}{\partial \theta^i} \frac{\partial G^b}{\partial \theta^j} \mathcal{I}^{ij}_{\tau}
\end{equation}
is the pushforward of the inverse Fisher metric by the map $G$ to give the induced Fisher metric on the data space itself.

We can now compare (\ref{Post Pred = Diffusion}) with (\ref{Diffusion Process}) and recognize that the posterior distribution, when pushed forward into the sample space by the map $G$, is the heat kernel of an associated convection-diffusion equation. In particular, (\ref{Post Pred = Diffusion}) can be read as the solution to a Fokker-Planck equation
\begin{equation} \label{Bayesian FP}
	\frac{dp_{\tau}}{d\tau} = \Delta p_{\tau} + \text{div}\left(p_{\tau} \text{grad}_{K_{\tau}} V_{\tau}\right)
\end{equation}
or equivalently, as describing a stochastic process
\begin{equation} \label{Bayesian SDE}
	dY^a_{\tau} = -\left(\text{grad}_{K_{\tau}} V_{\tau}\right)^a d\tau + \sqrt{2} (\theta_{\tau})^a_{b} dW^{b}_{\tau}.
\end{equation}
As discussed in \eqref{Bayesian Potential}, $V_{\tau}$ is a potential function on the sample space for which $\gamma_{\tau}$ is a gradient flow, and $\delta^{cd}(\theta_{\tau})^a_c (\theta_{\tau})^b_d = (K^{-1}_{\tau})^{ab}$.

Equations \eqref{Post Pred = Diffusion}, \eqref{Bayesian FP}, and \eqref{Bayesian SDE} complete an explicit mapping between a dynamical Bayesian inference scheme and an exact renormalization group flow which we can now represent schematically as the identification:
\begin{equation} \label{Bayes = ERG}
	(\mathcal{M}, \mathcal{I}, V) \leftrightarrow (S, K, V).
\end{equation}
The mapping \eqref{Bayes = ERG} provides a novel information theoretic interpretation for exact renormalization as the inverse process dual to a Bayesian inference procedure governed by \eqref{Dynamical Bayes}. In particular, we have realized our goal of demonstrating how performing the inverse of statistical inference, that is by discarding data rather than collecting data, implements an information theoretic renormalization scheme with emergent scale given by the distinguishability of probability distributions in model space. The role of this scale in coarse graining can be seen through the relationship between the ERG kernel, $K^{-1}_{\tau}$, and the Fisher information metric. 

\subsection{Comparison with Diffusion Learning} 

For clarity, it is helpful to compare the governing equations of Bayesian Renormalization and Dynamical Bayes with the forward and reverse diffusion processes introduced in Section \ref{sec: Diffusion}. From the diffusion learning perspective, one begins by running the data generating distribution through a forward diffusion process specified by an SDE of the form \eqref{Forward Diffusion}. This results in a more tractable distribution which can subsequently be used to generate samples from the data generating distribution by appealing to the associated reverse SDE \eqref{Reverse Diffusion}. To accomplish this step one must implement an algorithm which reconstructs the score functions; a task which can be formulated as a Bayesian learning problem. By contrast, our derivation of the Bayesian Renormalization scheme has ensued in exactly the opposite direction. We began with a Bayesian learning problem governed by the dynamical Bayesian inference equation \eqref{Dynamical Bayes}. Then, we constructed the reverse process (\emph{relative to Bayes}) by studying how the learned distribution changes when data is removed as opposed to incorporated. The result was a diffusion process governed by the SDE \eqref{Bayesian SDE}.

One should identify the forward diffusion, \eqref{Forward Diffusion}, with the Bayesian diffusion process, \eqref{Bayesian SDE}. Similarly, one should identify the reverse diffusion process, \eqref{Reverse Diffusion}, with the Dynamical Bayesian learning process, \eqref{Dynamical Bayes}. In summary, one way of interpreting Bayesian renormalization is as an information theoretic formulation of the observation that the reverse SDE associated with a diffusion process corresponds with a statistical learning problem. The advantage of this approach is that, opposed to a naive diffusion scheme as would be implemented by \eqref{Diffusion Process}, the Bayesian diffusion scheme intelligently discards information according to a hierarchy of importance as dictated by the Fisher metric. In this respect we hope that Bayesian Renormalization can inspire new approaches to information theoretic optimal diffusion learning while also providing insights into the underlying information theoretic character of both diffusion learning and generic renormalization.

To this end, although we have introduced Bayesian Renormalization through its relationship with Bayesian inference, we should stress that it motivates a methodology that can be undertaken without the need to first perform a Bayesian inference experiment. In particular, we can renormalize a family of probability distributions by diffusing it through the space of possible models with the diffusion kernel $\mathcal{N}(\gamma_{\tau}, \tau K^{-1}_{\tau})$; one can understand this procedure as information theoretic form of diffusion learning where Eqn. \eqref{Bayesian SDE} provides an explicit proposal for a forward diffusion process which coarse grains information in a hierarchy of importance as specified by the Fisher metric. In turn, the reverse SDE associated with \eqref{Bayesian SDE} through the identification \eqref{Reverse Diffusion} has an explicit interpretation as encoding the Bayesian learning of the data generating distribution in the time $T$ corresponding to the reintegration of removed data. Here, $K^{-1}_{\tau}$ should be chosen by first computing the Fisher information matrix of the model space of interest, and subsequently pushing this matrix forward in the sample space by selecting a function $G: \mathcal{M} \rightarrow S$. The potential function $V_{\tau}$ can be chosen arbitrarily, and will govern the mean tendency of the diffusion process. Thus, the Bayesian renormalization scheme corresponds to the choices of $G$ and $V_{\tau}$, with $\mathcal{I}$ being fixed by the system. This is significant because $\mathcal{I}$ is what encodes the information theoretic scale of the problem.

The identification of the mapping $G$ should be thought of as a form of data compression in which a deterministic model for the data realizations in terms of the parameters is postulated. Over the course of the Bayesian renormalization, there will be a flowing in these parameters resulting in the formulation of an effective model. This procedure is related to a variety of different data compression techniques such as the information bottleneck \cite{tishby2000information} and variational autoencoding \cite{Kingma:2013hel}. In this way, the Bayesian Renormalization scheme makes direct contact with previous insights into the relationship between model building/dimensional reduction for complex systems and the renormalization group such as \cite{Kline:2021ugg, Meshulam2018CoarseGF, Meshulam2018CoarsegrainingAH, PhysRevLett.126.240601}.

\section{Implementation of Bayesian Renormalization} \label{sec: app}

The preceding discussion has been largely abstract. Thus, in this penultimate section, we shall provide some insight into the overarching themes of our work through an explicit example\footnote{An example implementation of Fisher-motivated pruning at: \url{https://github.com/xand-stapleton/fisher_pruning}}. Rather than analyzing a Bayesian inference experiment, we apply the philosophy of Bayesian Renormalization directly to a Neural Network in order to stress the general applicability of our approach. In particular, we will demonstrate how Bayesian Renormalization can be used to systematically remove ``sloppy" parameters from a autoencoder network. 

\begin{table}[ht]
	\centering
	\begin{tabular}{|c|c|}
		\hline
		\textbf{Machine learner's approach} & \textbf{Bayesian's approach}\\ \hline
		Neural Network & Statistical Model\\
		Weights $w_i$ and biases $b_i$ & Model parameter $\theta_i$ \\
		Training Sample & Observed Data \\
    Trained weights/biases & Parameter terminal distribution $\theta^*$\\
    Minimization of loss function & Maximization of log-likelihood $\ln p_{Y|\Theta}(y | \theta)$\\
		\hline
	\end{tabular}
	\caption{Approximate dictionary mapping equivalent quantities in Bayesian statistics to those in the study of neural networks.}
  \label{tab:bayes_nn_dict}
\end{table}

Before delving into our experiment, it is instructive to reflect on the relationship between Bayesian inference and Machine Learning. To quote \cite{wasserman2004all}, ``Statisticians and computer scientists often use different language for the 
same thing." In this regard, it is helpful to provide a brief dictionary outlining the correspondence between Bayesian inference and the study of neural networks which can be found in Table \ref{tab:bayes_nn_dict}.
For a more extensive review of how Bayesian statistics is related to neural networks, see \cite{Jospin_2022}.

\subsection{Autoencoders}%
\label{sub:autoencoders}

An autoencoder is a type of (artificial) neural network trained using unsupervised learning which is used to create efficient encodings of a set of unlabelled input data. By definition, autoencoders are composed of two blocks: an encoder and a decoder; typically both blocks are trained simultaneously. 

Assuming the encoded and decoded message spaces are Euclidean, let
\begin{align}
	\chi &= \mathbb R^n& 
	\tilde \chi &= \mathbb R^m.
\end{align}
where $n$ denotes the dimension of the input data, and $m$ the dimension of the \textit{latent space} chosen when the network is initialized. For each data instance $y \in \chi$, we denote the encoded latent space representation as $\tilde y \in \tilde \chi$. Let $E_\theta$ be the encoder function, parameterized by a set of variables $\{\theta_i\}$ and mapping from the space of decoded (source) messages $\chi$ to the space of encoded messages $\tilde \chi$,
\begin{equation}
	E_\theta: \chi \to \tilde \chi.
\end{equation}
Similarly, $D_\phi$ is the decoding function, parameterized by a second set of variables $\{\phi_j\}$ and mapping from the space of encoded messages $\tilde \chi$ to the space of decoded messages $\chi$,
\begin{equation}
	D_\phi: \tilde \chi \to \chi.
\end{equation}

\begin{figure}[h]
	\centering
	    \begin{tikzpicture}[
        node/.style={draw, circle, minimum size=0.6cm},
        arrow/.style={-, shorten >=1pt, >=stealth, semithick}
    ]
        \foreach \x [count=\xi] in {1,...,4} {
            \node[node] (input\x) at (0, -\xi) {$x_{\x}$};
        }
				\node[above=0.1cm of input1] {Input};
        
        \foreach \x [count=\xi] in {1,...,2} {
            \node[node] (hidden\x) at (2.5, -\xi-1.0) {$h_{\x}$};
        }
				\node[above=0.1cm of hidden1] {Latent space};
        
        \foreach \x [count=\xi] in {1,...,4} {
            \node[node] (output\x) at (5, -\xi) {$y_{\x}$};
        }
				\node[above=0.1cm of output1] {Output};
        
        \foreach \x in {1,...,4} {
            \draw[arrow] (input\x.east) -- (hidden1.west);
            \draw[arrow] (input\x.east) -- (hidden2.west);
        }
        
        \foreach \x in {1,...,2} {
            \foreach \y in {1,...,4} {
                \draw[arrow] (hidden\x.east) -- (output\y.west);
            }
        }
    \end{tikzpicture}
	\caption{Highly degenerate model architecture with $n=4, m=2$. A simple MLP autoencoder with a pair of linear (dense) layers.}%
	\label{fig:autoencoder}
\end{figure}
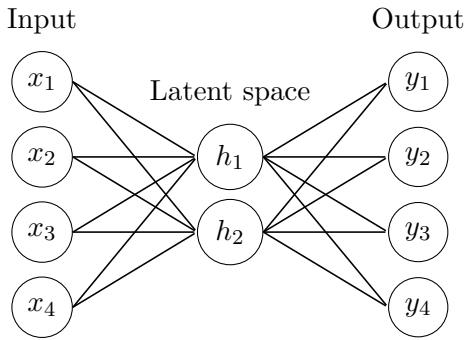

As a choice of autoencoder architecture, we consider a simple dense, (linear) multi-layer perceptron network with \textbf{no biases}, $n$ input neurons, and $m$ latent space neurons (where the number of output neurons is thus constrained also to $n$). Selecting the architecture as a multi-layer perceptron network is intentional: since each node is fully connected with an associated weight parameter, one may expect the network to have many weights that covary comparatively weakly with the output of the network. A schematic summary of the network architecture is presented in Figure \ref{fig:autoencoder}.

In the case of the aforementioned network, we choose the element-wise encoder and decoder layers to be
\begin{align}
	(E_\theta)_i &= \alpha_\text{Enc}(\theta_{ij} x^j)& 
	(D_\phi)_j &= \alpha_\text{Dec}(\phi_{ij} x^j),
\end{align}
where $\alpha_\text{Enc}, \alpha_\text{Dec}$ are the encoder and decoder activation functions\footnote{The activations are typically non-linear.}. Each parameter of the encoder and decoder networks are elements of the corresponding weight matrix $\theta_{ij}$ -- the machine learner may be more familiar with the usual form $\alpha(Wx + b)$ where $W$ is the typical weight matrix and $b$ is the corresponding bias vector. For all activations, we choose the \textit{Rectified Linear Unit (ReLU) activation} defined as
\begin{equation}
	\alpha(x) = \max(0, x) := 
	\begin{cases}
		x \text{ if } x > 0\\
		0 \text{ otherwise}
	\end{cases}
\end{equation}

Let $D \subset \chi$ denote the set of training data. Upon the introduction of each new item of training data, $y \in D$, the parameters of the autoencoder are updated\footnote{For simplicity, we choose the ADAM optimizer, but other choices such as stochastic gradient descent would work as well.} to minimize a given loss function $L(y \mid \theta,\phi)$. In the following numerical examples, we consider the typical \textit{mean squared error (MSE) loss} defined by
\begin{equation}
	L(y \mid \theta, \phi) = \frac{1}{n} \sum_{j=1}^{n} (y_j - D_{\phi}E_{\theta}(y)_j)^2 .
\end{equation}

\subsection{Information Shell Renormalization Scheme}%
\label{sub:pruning}
In the course of this note, we have stressed that a crucial question in any data science context is how we can arrive at an effective model for a given system using only the most relevant set of parameters. Indeed, it is typical for models to possess parameters that are comparatively less sensitive than others, and we have argued that these can be identified systematically via the Fisher matrix. As we have stressed in Table \ref{tab:bayes_nn_dict} a Neural Network, like our autoencoder, can be understood as a family of statistical models quantified in terms of the various parameters used to initialize its architecture. In our case, each set of parameters $(\theta,\phi)$ corresponds to a different autoencoder, with our best estimate of the ``true" or optimal autoencoder given by the value of these parameters which is realized after all of the training data have been utilized. We denote these optimal parameters by $(\theta^*,\phi^*)$. Appealing again to Table \ref{tab:bayes_nn_dict}, we can give a more decidedly probabilistic viewpoint on the family of autoencoders by regarding each autoencoder $(\theta,\phi)$ as initializing a probability distribution over data of the form
\begin{equation} \label{Autoencoder Probability Model}
	p_{Y \mid \Theta,\Phi}(y \mid \theta,\phi) \propto \text{exp}\bigg(-L(y \mid \theta,\phi)\bigg).
\end{equation}
Then, the minimization of the loss can equivalently be regarded as a maximization of the log-likelihood specified in \eqref{Autoencoder Probability Model}.

Regarding the set of all possible autoencoders as a family of statistical models, we can employ the machinery developed in Section \ref{sec: infogeo} to endow this space with an information geometry. In particular, the Fisher matrix of the \textbf{encoder layer} may be calculated explicitly from the \textbf{trained model} using the approach outlined in \cite{george_nngeometry}. During this analysis the trained parameters of the decoder are left untouched -- this allows comparison between pruning methods while eliminating inter-layer interactions. We denote by $\mathcal{I}_{ij}$ the components of the resulting Fisher matrix. Because it is computed after training one should regard this as the Fisher metric evaluated at the point $(\theta^*,\phi^*)$.

As we have discussed, the eigenvalues of the Fisher metric serve as a proxy for the relevance of the various parameters which are included in the model class. From a physical perspective, one may think of each parameter $\theta_i$ as a mode of a particular theory, and of its associated diagonal element in the Fisher metric $\mathcal{I}_{ii}$ as corresponding to its characteristic ``length scale". Parameters with small Fisher diagonals vary weakly under the acquisition of new data and therefore require either extremely fine measurements or very large amounts of data in order to substantiate even an incremental change in their trained value. For this reason, we can think of these parameters as changing only on very fine length scales in the space of models. Conversely, parameters with larger Fisher values vary strongly with the model and therefore possess features that can be distinguished more broadly with less or less fine-tuned data.  

In light of these observations, we now propose the most simple-minded form of Bayesian renormalization. We introduce a sliding parameter, $\Lambda$, and divide the set of parameters into two subsets:
\begin{equation}
	\Theta^{>}_{\Lambda} \equiv \{\theta_i \; | \; \mathcal{I}_{ii} > \Lambda \}, \qquad \Theta^{<}_{\Lambda} \equiv \{\theta_i \; | \; \mathcal{I}_{ii} \leq \Lambda\}.
\end{equation}
The set $\Theta^{>}_{\Lambda}$ consists of those parameters that covary with the model output on ``length" scales greater than $\Lambda$. We refer to these parameters equivalently as relevant/stiff parameters relative to the scale $\Lambda$ for obvious reasons. Conversely, the set $\Theta^{<}_{\Lambda}$ consists of those parameters that covary with the model output on ``length" scales below $\Lambda$. We refer to these parameters equivalently as irrelevant/sloppy relative to the scale $\Lambda$. The renormalization procedure we shall consider is simply to remove the parameters in $\Theta^{<}_{\Lambda}$. This realizes a reduced/renormalized model that depends only on the parameters $\Theta^{>}_{\Lambda}$. One should think of this as akin to a hard cutoff regularization scheme in a Quantum Field Theory in which modes that depend on physics at length scales smaller than a given cutoff $\Lambda$ are systematically removed. 

\subsection{Experimental results}%
\label{sub:experimental_results}

\begin{table}[ht]
        \centering
        \begin{tabular}{@{}|c|c|@{}}
            \hline
            \textbf{Model property} & \textbf{Value}\\ \hline
            Input (Output) dimension & $784$ \\
            Latent space dimension & $20$\\
            Number of encoder parameters (pre-prune) & $15680$\\
            Mean diagonal Fisher value & $0.7148$\\
            Optimizer & ADAM\\
            ADAM $(\alpha, \beta_1, \beta_2, \epsilon)$ & $(0.001, 0.9, 0.999, 10^{-8})$ \\
            Cutoff $\Lambda$ & $(0, 0.125, 0.25, ..., 2.375)$\\
            \hline
        \end{tabular}
        \caption{Values of configurable properties of the autoencoder network.}
        \label{tab:network_params}
\end{table}
For our experiment, we begin by training the autoencoder described in Section \ref{sub:autoencoders} on the MNIST dataset. Recall that MNIST is a series of approximately 50 thousand training samples and 10 thousand testing samples consisting of $28 \times 28$ pixel (784 dimensional) images depicting handwritten digits 0 through 9. The network and optimizer\footnote{ADAM parameters are defined in \cite{kingma2017adam}.} settings are outlined in Table \ref{tab:network_params}. 

\begin{figure}[H]
    \centering

    \begin{subfigure}{0.45\textwidth}
        \includegraphics[width=\linewidth]{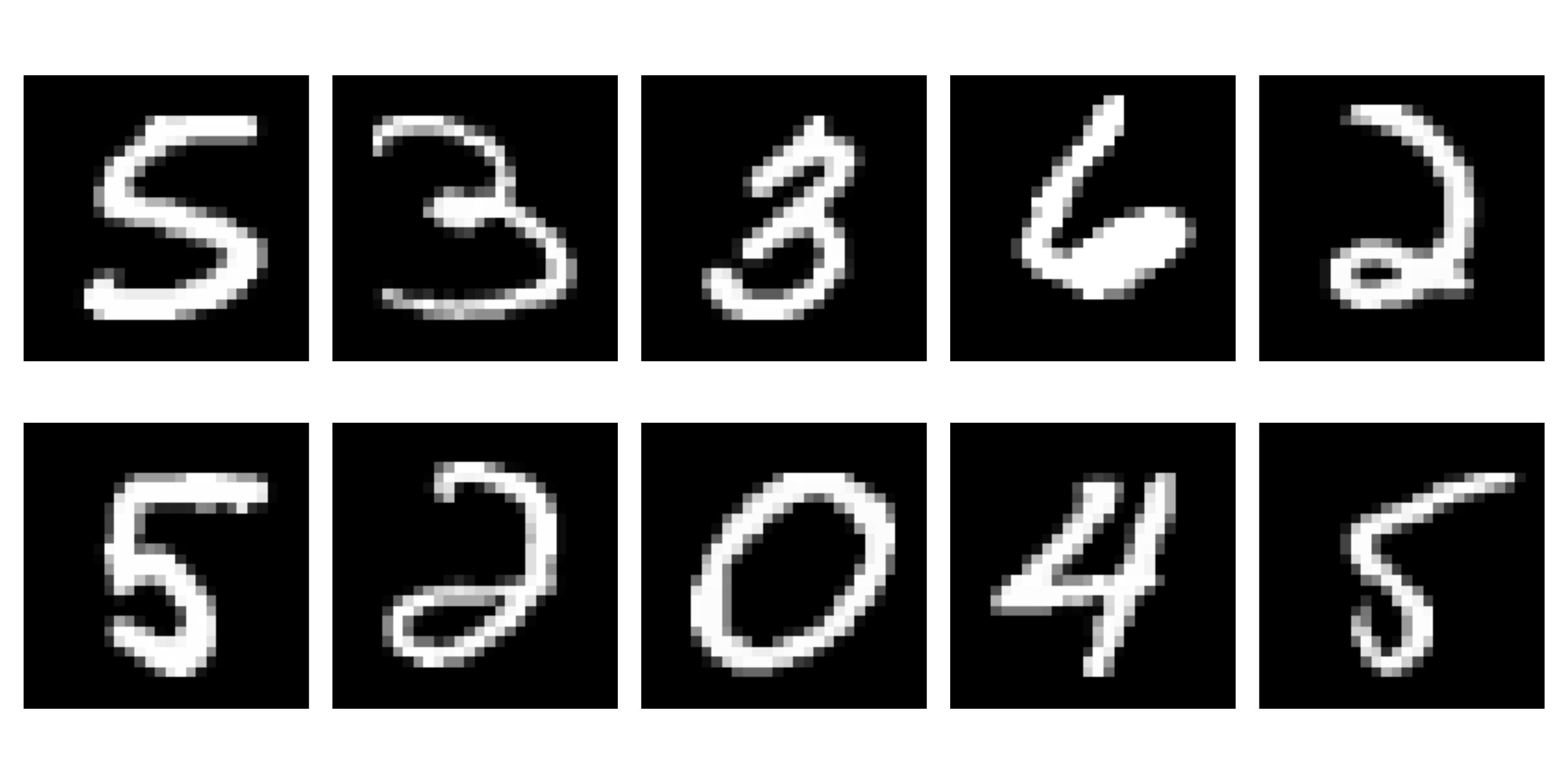}
        \caption{Source images. Input to autoencoder.\newline}
    \end{subfigure}\hfil
    \begin{subfigure}{0.45\textwidth}
        \includegraphics[width=\linewidth]{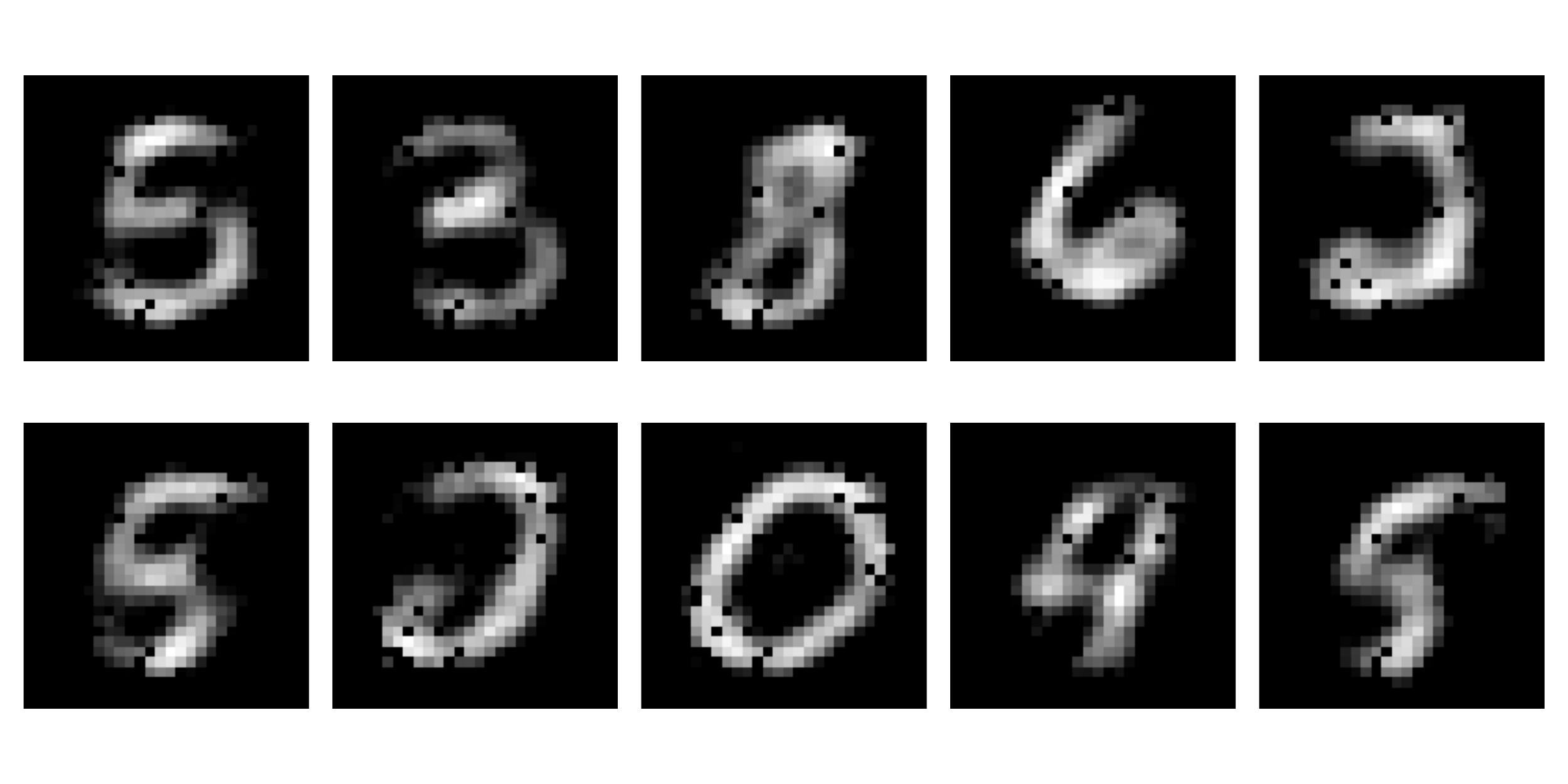}
				\caption{Cutoff $\Lambda=0$. 15680 parameters. Unpruned autoencoder.}
    \end{subfigure}\hfil

    \begin{subfigure}{0.45\textwidth}
        \includegraphics[width=\linewidth]{plots/autoencoded_stiff_cutoff_0.125_rem_10168_of_15680_params.pdf}
				\caption{Cutoff $\Lambda=0.125$. 10168 parameters removed (64.8\%).}
    \end{subfigure}\hfil
    \begin{subfigure}{0.45\textwidth}
        \includegraphics[width=\linewidth]{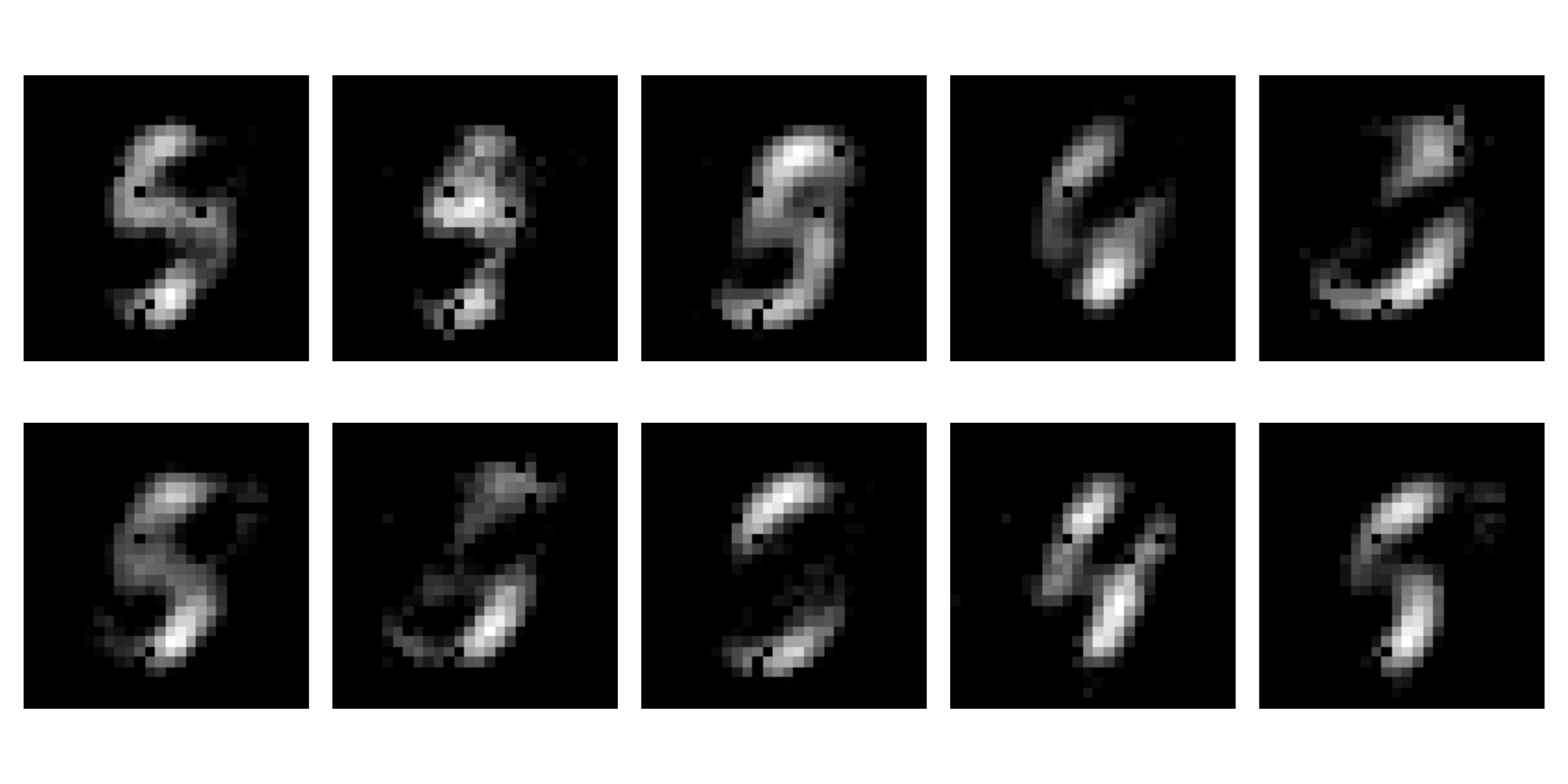}
				\caption{Cutoff $\Lambda=1.25$. 13879 parameters removed (82.1\%).}
    \end{subfigure}\hfil

    \begin{subfigure}{0.45\textwidth}
        \includegraphics[width=\linewidth]{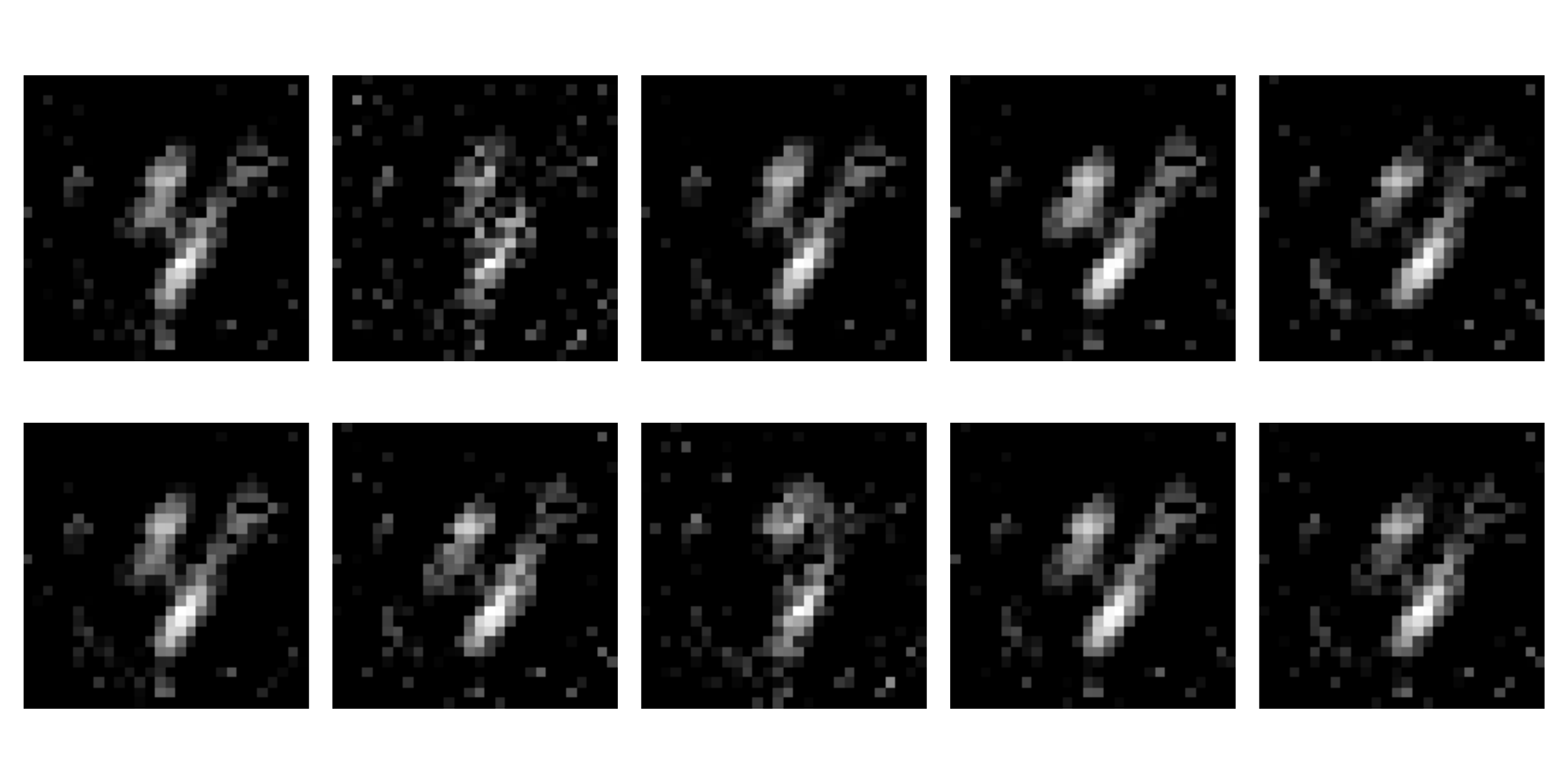}
				\caption{Cutoff $\Lambda=1.75$. 15386 parameters removed (98.1\%).}
    \end{subfigure}\hfil
    \begin{subfigure}{0.45\textwidth}
        \includegraphics[width=\linewidth]{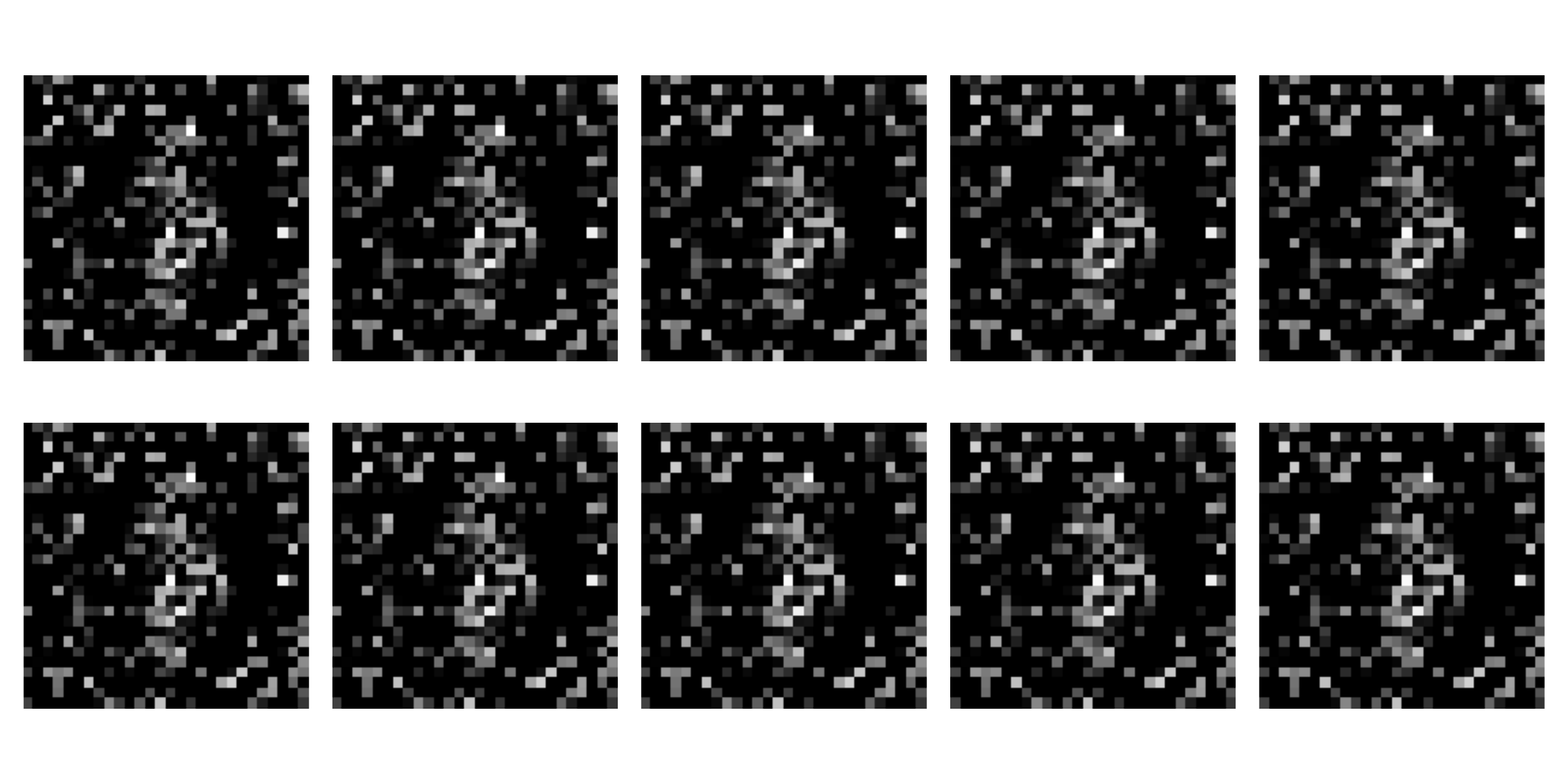}
				\caption{Cutoff $\Lambda=2.375$. All (15680) parameters removed (100\%).}
    \end{subfigure}\hfil

    \caption{Fisher-motivated pruning with variable cutoffs $\Lambda$.}
		\label{fig:fisher_pruning}
\end{figure}

\begin{figure}
    \centering

    \begin{subfigure}{0.45\textwidth}
        \includegraphics[width=\linewidth]{plots/original_samples.pdf}
        \caption{Source images. Input to autoencoder.\newline}
    \end{subfigure}\hfil
    \begin{subfigure}{0.45\textwidth}
        \includegraphics[width=\linewidth]{plots/autoencoded_stiff_cutoff_0.125_rem_10168_of_15680_params.pdf}
				\caption{15680 parameters. Unpruned autoencoder. Equivalent to cutoff $\Lambda=0$.}
    \end{subfigure}\hfil

    \begin{subfigure}{0.45\textwidth}
        \includegraphics[width=\linewidth]{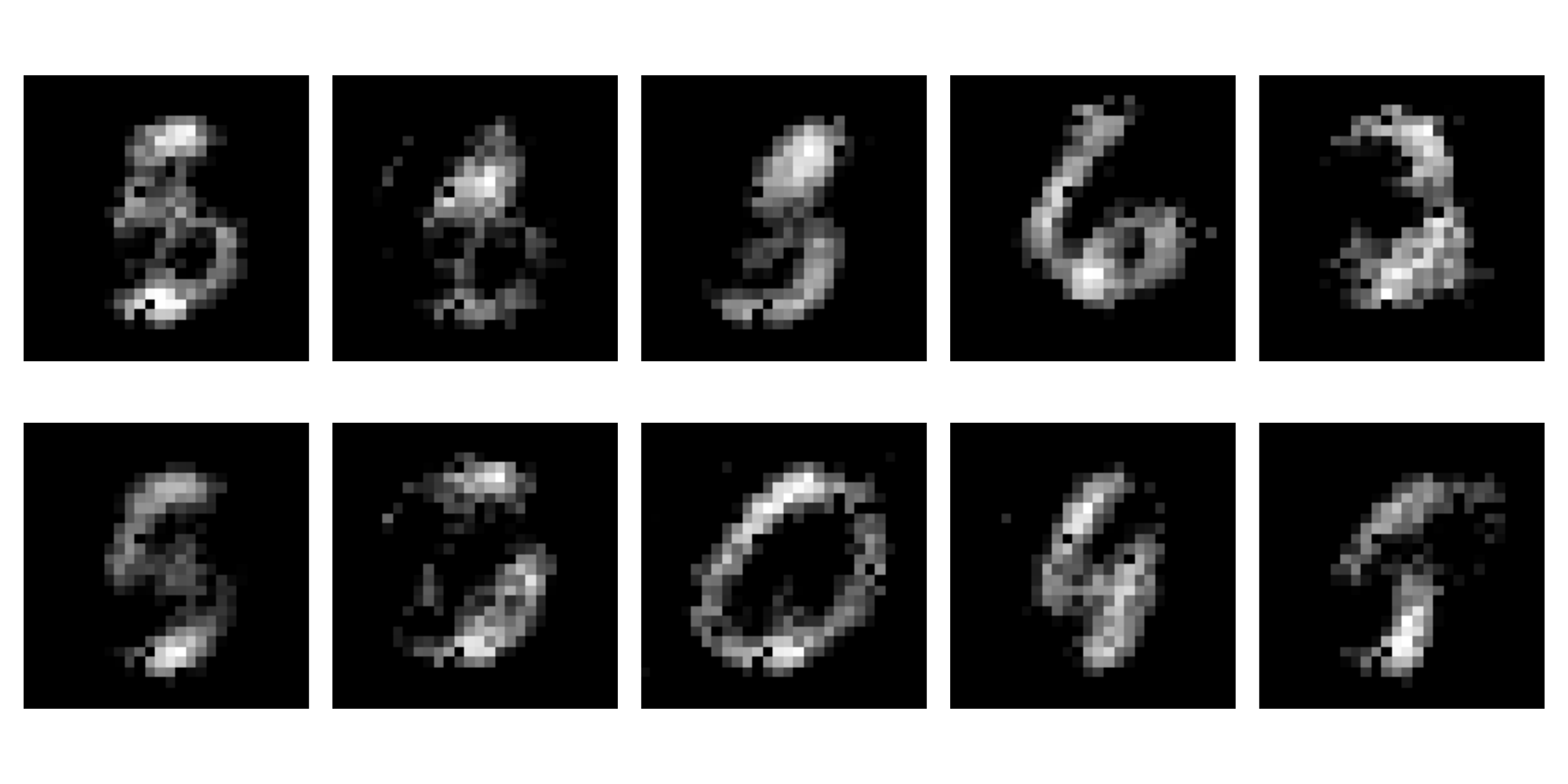}
				\caption{10168 parameters removed (64.8\%). Equivalent to cutoff $\Lambda=0.125$.}
    \end{subfigure}\hfil
    \begin{subfigure}{0.45\textwidth}
        \includegraphics[width=\linewidth]{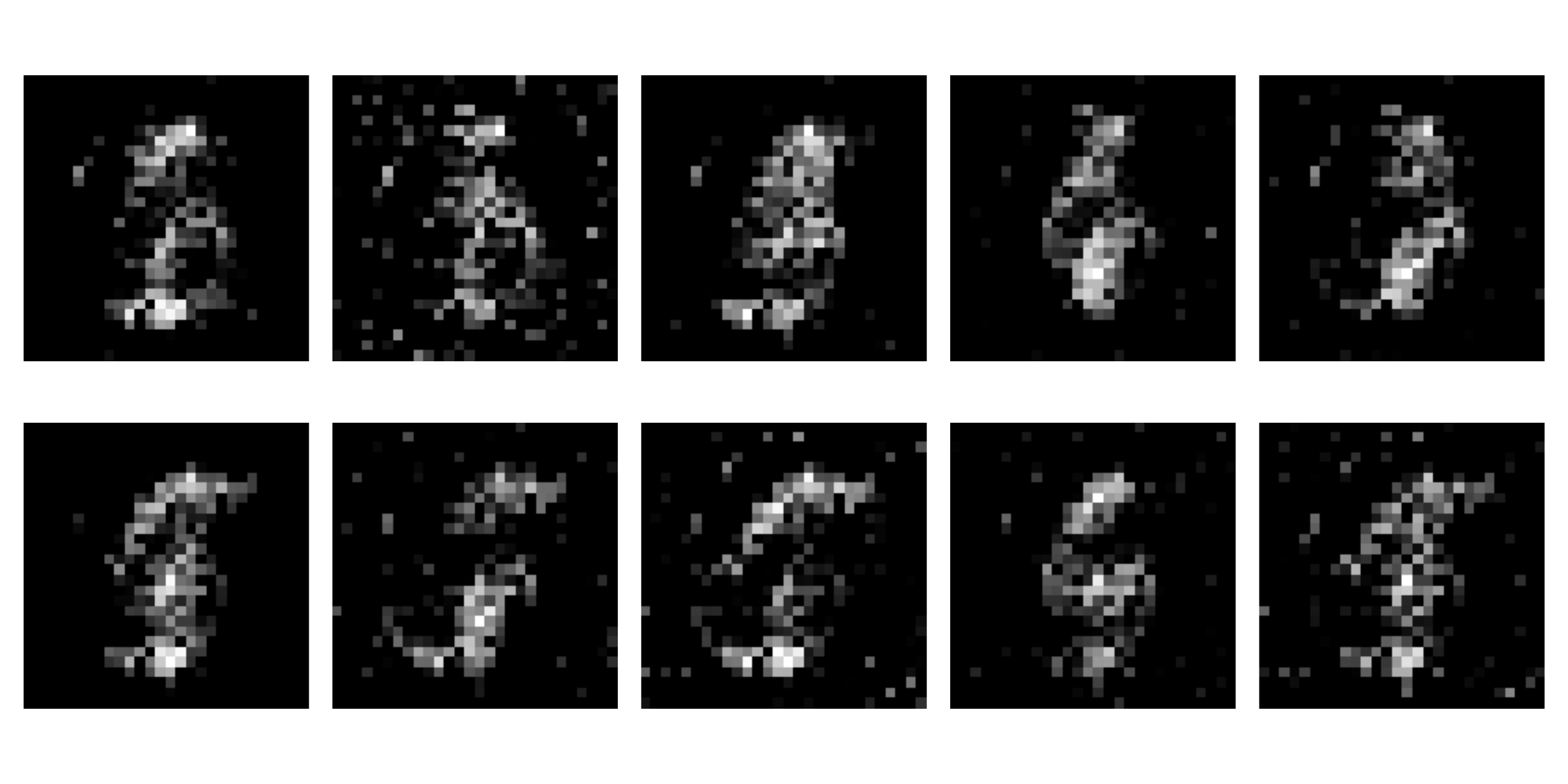}
				\caption{13879 parameters removed (82.1\%). Equivalent to cutoff $\Lambda=1.25$.}
    \end{subfigure}\hfil
    \begin{subfigure}{0.45\textwidth}
        \includegraphics[width=\linewidth]{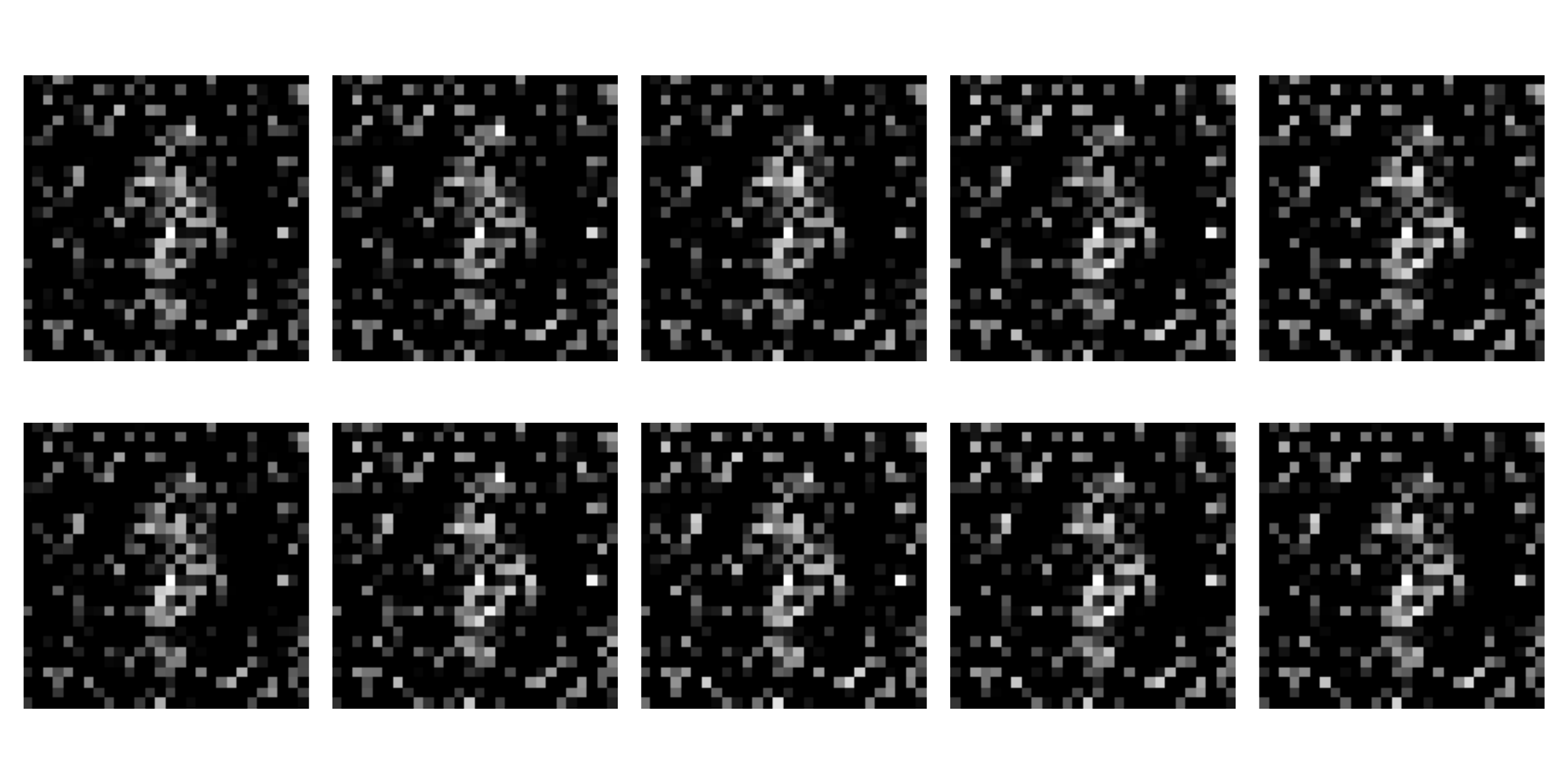}
				\caption{15386 parameters removed (98.1\%). Equivalent to cutoff $\Lambda=1.75$.}
    \end{subfigure}\hfil
    \begin{subfigure}{0.45\textwidth}
        \includegraphics[width=\linewidth]{plots/autoencoded_stiff_cutoff_2.375_rem_15680_of_15680_params.pdf}
				\caption{All (15680) parameters removed (100\%). Maximal information loss. Equivalent to cutoff $\Lambda=2.375$.}
    \end{subfigure}\hfil

		\caption{Randomly pruned network with equivalent numbers of removed parameters to Figure \ref{fig:fisher_pruning}.}
		\label{fig:uniform_pruning}
\end{figure}

After having trained the model, we perform the Bayesian renormalization scheme outlined above. The result of this procedure at various values of the cutoff parameter $\Lambda$ can be seen in Figure \ref{fig:fisher_pruning}. To illustrate the sense in which the Bayesian RG scheme more adequately coarse grains the model with respect to the information contained in its various parameters we have also provided two alternative schemes. In the first scheme, demonstrated in Figure \ref{fig:uniform_pruning}, the same number of parameters are removed in each iteration as in the Fisher-inspired scheme, but using a uniform distribution over the parameters of the model. The second alternative scheme is a popular form of pruning for neural networks which we refer to as ``magnitude inspired pruning". In the magnitude inspired approach a fixed number of parameters are removed in each iteration corresponding to the remaining parameters with the smallest absolute value, hence the name. 

\begin{figure}[H]
	\centering
\includegraphics[width=0.6\linewidth]{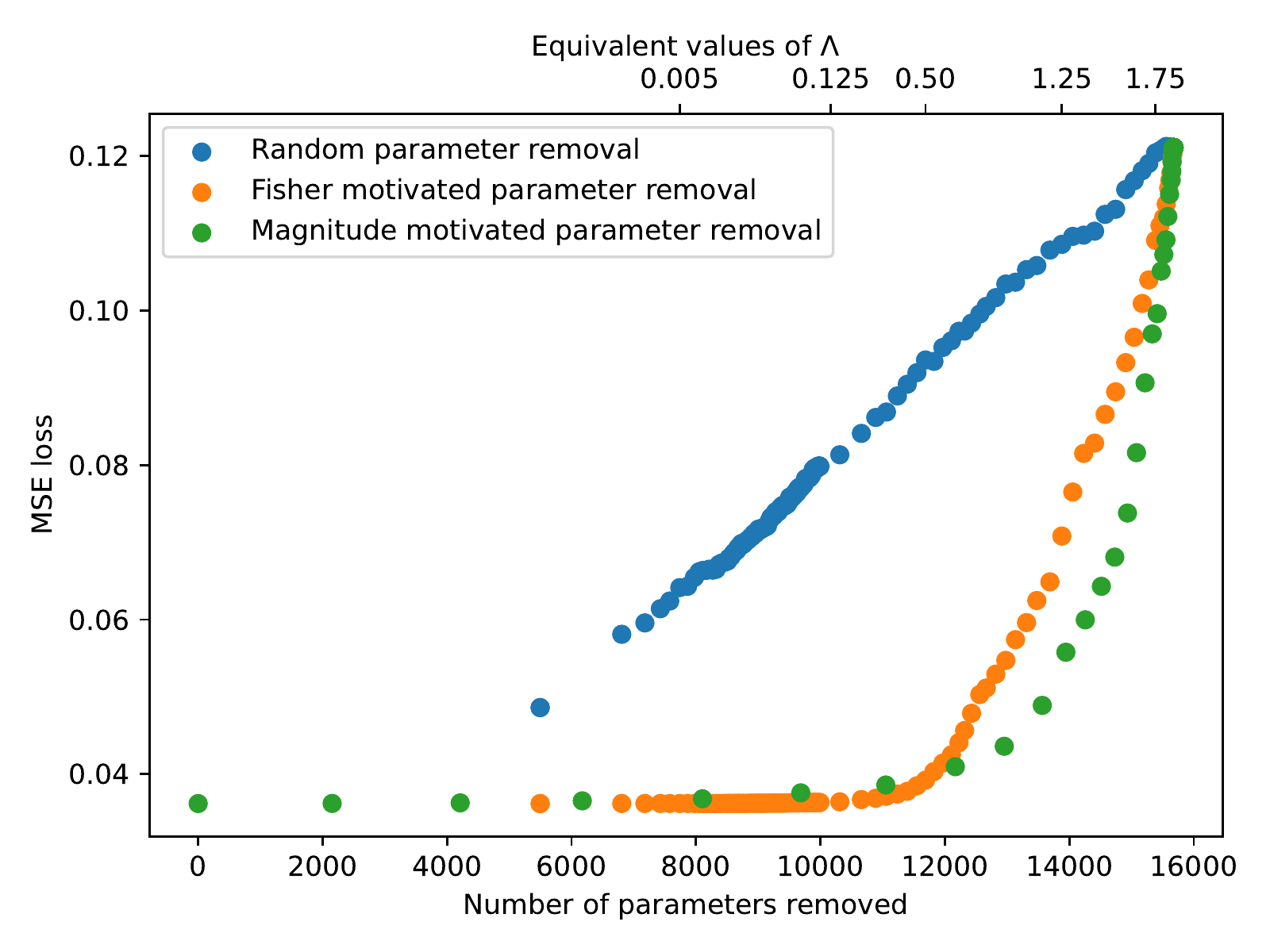}
	\caption{Mean square error loss between the true distribution and output of the pruned networks.}%
	\label{fig:losses_uniform_fisher}
\end{figure}

\begin{figure}[H]
	\centering
	\includegraphics[width=0.6\linewidth]{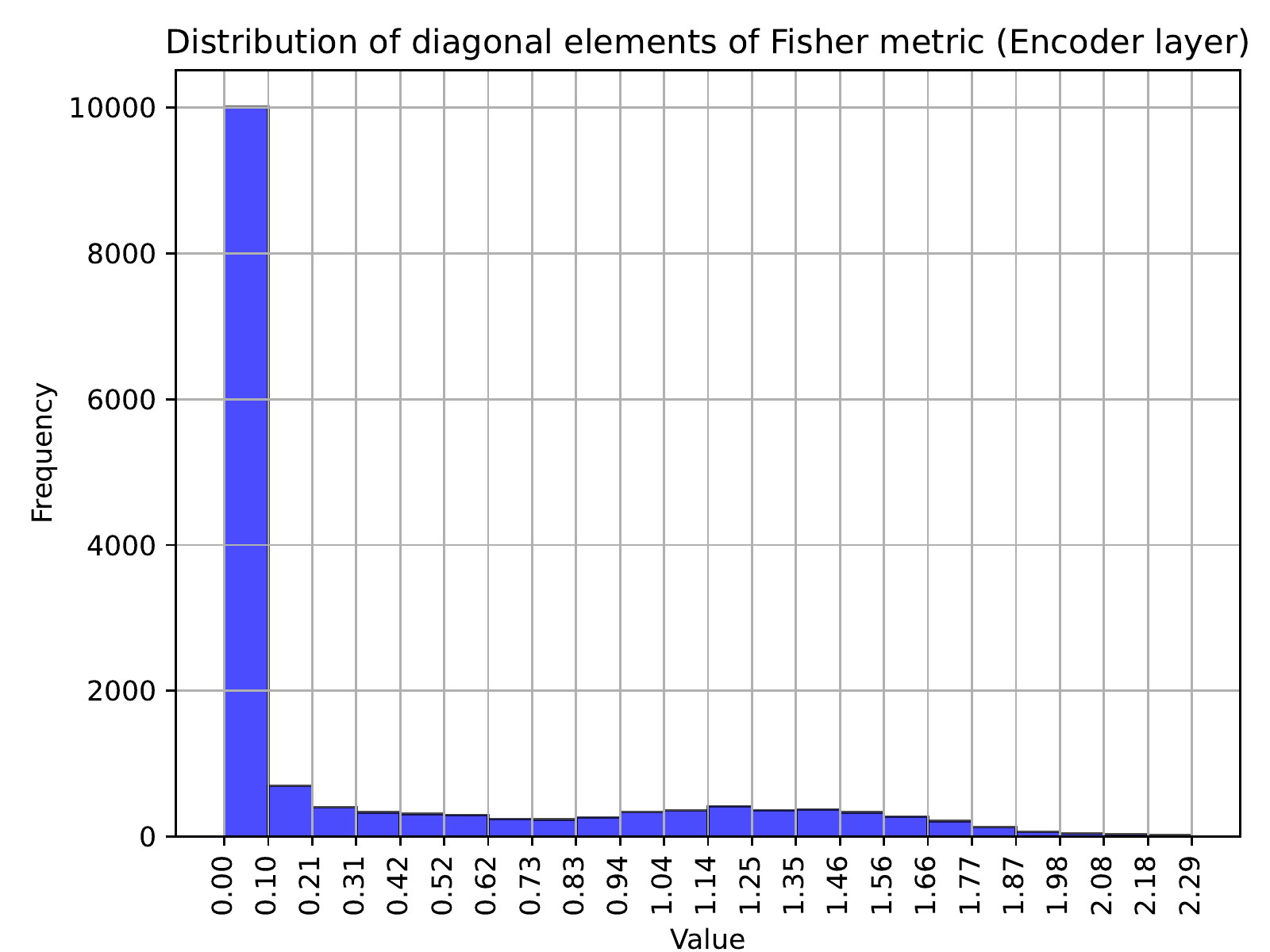}
  \caption{Histogram showing the distribution of diagonal Fisher matrix elements.}%
	\label{fig:fisher_histogram}
\end{figure}

As one can see in Figure \ref{fig:fisher_pruning}, the autoencoder remains remarkably effective even once more than half of its parameters have been removed, provided these parameters are removed using the Bayesian scheme. To understand this result it is instructive to look at the distribution of (diagonal) Fisher elements associated with the encoder -- see Figure \ref{fig:fisher_histogram}. Here one can see that there is a large gap in the spectrum at $\Lambda \sim 0.1$ -- we regard this gap as identifying the genuinely ``sloppy" parameters in the model. In Figure \ref{fig:losses_uniform_fisher} we explore the loss landscape as a function of the number of removed parameters, comparing the Fisher-motivated scheme with random parameter removal and magnitude inspired pruning. In regards to this comparison there are two important points to be made. Firstly, the Fisher-motivated scheme produces minimal losses up to the aforementioned cutoff at $\Lambda \sim 0.1$. Shortly thereafter, the losses of the Fisher motivated pruning scheme exceed those of the magnitude scheme. This is expected; for $\Lambda > 0.1$ the Fisher scheme begins to remove parameters that are relevant to the performance of the model. The loss responds strongly to the removal of these paramters. In other words, Figure \ref{fig:losses_uniform_fisher} demonstrates that the Fisher scheme effectively distinguishes both irrelevant and relevant parameters.

A different way of understanding the preceding discussion serves to contextualize the relationship between RG and statistical learning. One can imaging performing the pruning procedure in reverse, whereby one incorporates new paramters one at a time in order to improve the model's performance. From this perspective, Figure \ref{fig:losses_uniform_fisher} demonstrates that the Fisher scheme most rapidly reaches minimal losses among the three approaches here considered. 

One can summarize the preceding discussion by observing the distinctive shape of the loss curve produced by the Fisher pruning scheme. Up to the aforementioned ``scale" of $\Lambda \sim 0.1$ the loss is essentially stable to the pruning of parameters. Thereafter, the removal of subsequent parameters results in noticeable decreases in the performance of the model. The result is a `hockey-stick' shaped curve: for $\Lambda \lesssim 0.1$ the pruning removes irrelevant parameters that covary weakly with the output of the model, and thus only marginally affect the loss; likewise for $\Lambda > 0.1$ the pruning begins to remove `relevant' parameters that covary heavily with the output of the model. In other words, the Bayesian RG scheme has identified a characteristic ``scale" in the model and has dictated that the inclusion of parameters which are irrelevant with respect to that scale may be excluded from the model without any change in performance -- this is precisely the goal of an RG flow. It is instructive to recognize that the notion of scale which is relevant in this instantiation of RG has a direct relationship with the number of parameters included in the model. 

\section{Discussion} \label{sec: Discussion}

In this note we have outlined a new perspective on renormalization that is fully information theoretic in nature. As a consequence, this approach to renormalization is amenable to arbitrary probability models, not just those which possess a physical interpretation. The major insight we have presented is that the Fisher metric should be interpreted as defining a correlation length in the space of models which defines an emergent scale through the distinguishability of probability distributions. From this perspective, a UV cutoff in a Bayesian renormalization scheme can be understood as fixing the maximum number of possible measurements that could be made about a system, thereby bounding the precision with which one can access information about the data generating model in an inference experiment. This perspective on renormalization is consistent with the more familiar physical picture where the amount of data that can be collected about a theory is bounded by the energy scales which can be probed experimentally. Hence, again, the UV cut-off dictates the set of possible independent measurements. More rigorously, this observation can be quantified through the observation that the KL-divergence is a ERG monotone \cite{cotler2022renormalization}, or the fact that even in RG schemes with a physical scale the ERG kernel can be identified with the Fisher metric on the space of theories \cite{2303.04082v1}. 

The information theoretic approach to renormalization allows for a very satisfying connection to be made between renormalization and techniques from data science such as model selection, data compression, and data generation. Among the most important insights that arise from using the Fisher metric to define the emergent RG scale is that ``high energy modes" are naturally identified with ``sloppy" parameters which are systematically discarded to formulate ``low energy" effective theories that depend only on ``strict" parameters. In this respect, we regard Bayesian Renormalization as an information geometrically inspired coarse-graining scheme. A related perspective on the relationship between renormalization and data compression has been studied through the so called  {\it{information bottleneck}}, for a representative sample of papers on the topic see \cite{Meshulam2018CoarseGF, Meshulam2018CoarsegrainingAH, Kline:2021ugg, PhysRevLett.126.240601}. The basic idea here is that the information bottleneck identifies a set of low dimensional effective degrees of freedom which efficiently encode the data contained in an otherwise high dimensional space of data realizations. Moving from the original degrees of freedom to the effective degrees of freedom involves a stochastic mapping (conditional probability distribution) which may be interpreted as a form of coarse graining based renormalization. In addition to the connection with data compression, the framing of Bayesian renormalization as a stochastic diffusion process allows one to interpret it as a refinement on the influential diffusion learning paradigm introduced in \cite{sohl2015deep}. The duality between renormalization and statistical learning provides new context for the usefulness of score based generative algorithms for inverting diffusion processes. 

At the moment the connections we have addressed are largely conceptual, but in future work we hope to demonstrate how the understanding of Bayesian renormalization can contribute to new automated techniques for data compression and data generation. As a first pass at this problem, we have explored a simple implementation of Bayesian renormalization to a Neural Network in Section \ref{sec: app} to illustrate some of its most salient and useful features. The most crucial insight from this example is that Fisher inspired renormalization can systematically discard degrees of freedom in a hierarchy of importance relative to the model in question. The specific approach we used is a one-to-one adaptation of Wilson's momentum shell renormalization scheme in which momentum shells are replaced by regions of fixed radius in the space of models according to the measure of distance defined by the Fisher metric. No assumptions about the structure or symmetries of the data or model were required in order to perform the aforementioned renormalization. Rather, the structure of the system is learned and subsequently encoded in the hierarchy of scales communicated by the Fisher metric.

Although we have concentrated primarily on the value of the information theoretic character of Bayesian renormalization in data science contexts, let us close by noting that this approach to renormalization may also provide new insights in physics contexts as well. For starters, the information theoretic approach of Bayesian renormalization makes it an ideal tool for identifying and quantifying the precise information that is lost under an RG flow. In this way, one should be able to use Bayesian renormalization in order to interpret and construct RG monotonicity theorems \cite{zamolodchikov1986irreversibility, alvarez1999geometric, myers2010seeing, casini2007c, casini2015mutual, casini2017markov}. On a different note, a modern perspective on renormalization would not be complete without including Entanglement/Holographic Renormalization: \cite{swingle2012entanglement, nozaki2012holographic, mollabashi2014holographic, leigh2014holographic, Leigh:2014qca, evenbly2015tensor, furuya2020real, goldman2023exact}. It is a challenge for future work to bring together Bayesian renormalization described in this paper with the holographic description of renormalization that has been developed in the above works. Of relevance to this is the relationship between canonical energy and Fisher metric in holographic context: \cite{Lashkari:2015dia, lashkari2016canonical, Banerjee_2018, Faulkner_2017, Erdmenger_2020}. Finally, the different ways that energy scales with entropy in a physical system tell us about the effectiveness of the usual momentum-based renormalization. An interesting question that one might ask is whether there is a different way of performing renormalization that more appropriately coarse grains information. In particular, large momentum shells may not be the right ``sloppy parameters" for a gravitation theory. Recall that the scaling of entropy with energy is different in gravity from that of local quantum field theories. This is a key ingredient that allows the holographic property of gravity. The perspective presented here is that the energy cut-off is really an information cut-off. As such this information theoretic perspective suggests that one might consider a different cut-off scheme for gravity than one uses for QFT.\footnote{A manifestation of this idea can be seen in \cite{Freidel:2022ryr,Freidel:2023ytq}, where the authors identified a relationship between the vacuum energy and entropy, and subsequently proposed a new approach to the cosmological constant problem in which the regulator scheme constrains the number of gravitational degrees of freedom by evoking the holographic principle.} This is the power of the Bayesian Renormalization principle: it automatically encodes the appropriate designation of relevant and irrelevant degrees of freedom through the Fisher metric and thus ensures that the degrees of freedom that are ``integrated out of the model" correspond precisely with the sloppy parameters, whatever they may be.

\section*{Acknowledgments}

We thank Jonathan Heckman for collaboration on dynamical Bayes which led to many of the ideas in this paper. We are also grateful for comments from Semon Rezchikov and Miranda Cheng on related work which was presented during Stringdata2022, and to Leenoy Meshulam, Adam Kline and Michael Abbott for helpful conversations about the intersection between renormalization and data science at the APS March Meeting 2023. Finally, we would like to thank Samuel Goldman and Robert Leigh for enlightening discussions on Exact
Renormalization and its relationship with Entanglement Renormalization. DSB and AGS acknowledge support from Pierre Andurand over the course of this
research. MSK is supported through the Physics department at the University of Illinois at
Urbana-Champaign.

\bibliographystyle{JHEP} 
\bibliography{arXiv}


\end{document}